\documentclass[a4paper,twocolumn,11pt9,unpublished]{quantumarticle}
\pdfoutput=1
\usepackage[utf8]{inputenc}
\usepackage[T1]{fontenc}
\usepackage{amsmath,amssymb,amsfonts,amsthm}
\usepackage{graphicx}
\usepackage{textcomp}
\usepackage{caption}
\usepackage{subcaption}
\usepackage[sort, compress, noadjust]{cite}
\usepackage{hyperref}
\usepackage{booktabs}
\usepackage{url}
\usepackage{flushend}
\usepackage{multirow}
\usepackage{array}
\usepackage{tabularx}
\usepackage{appendix}

\begin{document}

\newcommand{\todo}[1]{\noindent{\textbf{\textcolor{red}{[TODO: #1]}}}}
\newcommand{\eg}{e.\,g., }
\newcommand{\Eg}{E.\,g., }
\newcommand{\ie}{i.\,e.\@ }
\newcommand{\Ie}{I.\,e., }

\renewcommand{\appendixpagename}{Appendix}

\def\sectionautorefname{Sec.}
\def\subsectionautorefname{Sec.}
\def\subsubsectionautorefname{Sec.}
\def\figureautorefname{Fig.}
\def\tableautorefname{Tab.}

\makeatletter
\renewcommand{\fnum@figure}{Fig.\ \thefigure}
\renewcommand{\fnum@table}{Tab.\ \thetable}
\makeatother

\title{Shuttling Compiler for Trapped-Ion Quantum Computers Based on Fine-Tuned Large Language Models}

\author{Fabian Kreppel}
\affiliation{Institute of Computer Science, Johannes Gutenberg University, Staudingerweg 9, 55128 Mainz, Germany}
\email{f.kreppel@uni-mainz.de}
\orcid{0000-0002-8523-2112}
\author{Reza Salkhordeh}
\affiliation{Institute of Computer Science, Johannes Gutenberg University, Staudingerweg 9, 55128 Mainz, Germany}
\email{rsalkhor@uni-mainz.de}
\orcid{0000-0003-3786-7102}
\author{Ferdinand Schmidt-Kaler}
\affiliation{Institute of Physics, Johannes Gutenberg University, Staudingerweg 7, 55128 Mainz, Germany}
\email{fsk@uni-mainz.de}
\orcid{0000-0002-5697-2568}
\author{André Brinkmann}
\affiliation{Department of Computer Science, Saarland University, Saarland Informatics Campus, E1 1, 66123 Saarbrücken, Germany}
\email{andre.brinkmann@uni-saarland.de}
\orcid{0000-0003-3083-2775}

\maketitle

\begin{abstract}
In trapped-ion quantum computers, qubits must be shuttled between segments to interact. The routing logic that schedules these movements is written by hand for every new trap architecture. We present shuttling compilers based on five large language models (LLMs). Each LLM is fine-tuned on shuttling schedules produced by hand-coded heuristics for linear and branched one-dimensional trap architectures. We investigate how the shuttling operation counts of their schedules compare with those of the heuristics and how far they generalize to unseen architectures. For circuits of up to 16 qubits, the fine-tuned LLMs generate valid schedules on both training architectures, more often the fewer qubits a circuit has. In 12\,\% of the compilations yielding a schedule, the best of ten runs needs up to 21\,\% fewer operations than the heuristic baselines, after a rule-based post-processing step. A single run of one fine-tuned LLM produces a valid shuttling schedule for a previously unseen four-way branched architecture. This is preliminary evidence of cross-architecture generalization. On two other unseen architectures no LLM produces a valid schedule. Thus, LLM-learned shuttling compilation is feasible, and we show how far it currently reaches.
\end{abstract}

\section{Introduction}
\label{sec:introduction}

Compilation for shuttling-based trapped-ion quantum computers \cite{KielpinskiMW02, KaushalLSHPSBMSP20, PinoDFGMABFHMRN21, MuraliDBM20, MosesB+23} has so far relied almost exclusively on heuristics designed by hand, each tailored to one specific trap architecture. As the field moves from linear segmented ion traps to branched, racetrack \cite{MosesB+23}, and junction-based layouts \cite{Ransford+26}, the number of architectures for which such a heuristic has to be written is growing. In this work, we present the first shuttling compilers based on large language models (LLMs), which learn the routing strategy from data. We investigate how their shuttling operation counts compare with those of the heuristics and how far they transfer to unseen trap architectures.

The relevance of shuttling compilation is dictated by the hardware. In current segmented microchip ion-trap experiments, individual shuttling primitives take on the order of tens of microseconds per operation, comparable to or longer than the gate operations themselves \cite{HilderPOSOLRMSP22}. Shuttling therefore consumes a substantial part of the qubit coherence budget: it accounts for most of the runtime of a gate sequence and leaves the ions motionally excited, an effect that introduces dephasing in the subsequent gates and errors in the readout \cite{HilderPOSOLRMSP22}. Reducing the number of shuttling operations is one of the most direct ways to improve end-to-end fidelity on such devices.

In shuttling-based trapped-ion quantum computers, information is encoded in the internal states of individual ions. Accordingly, we use the term \emph{qubit} to refer to an ion throughout this paper. Qubits are confined in segmented microchip traps \cite{SchulzPSS06}, where each segment generates a localized potential well. Adjusting the electrode voltages reshapes these wells and thereby \emph{shuttles} qubits from one segment to the next. Segments are either \emph{qubit segments}, where qubits are held, or \emph{gate segments}. Gate operations are driven by laser or microwave fields and can be applied only to qubits located in a gate segment.

We restrict ourselves to a single gate segment throughout this work. This isolates the routing problem from the separate question of how to schedule gates that could run in parallel in different segments. It also leaves more operations for the compiler to plan than a trap with several gate segments would, since the qubits of every gate have to be brought to that single segment. The restriction reflects current hardware as well, where the number of gate segments is small \cite{KaushalLSHPSBMSP20, SchmaleTBPKDOWB22}, although devices with several gate segments have recently been realized \cite{MosesB+23, Ransford+26}.

Most current architectures arrange trap segments in a linear array \cite{KaushalLSHPSBMSP20, PinoDFGMABFHMRN21, BowlerGLTHJHLW12, LeeJPJKC21} or in a racetrack loop \cite{MosesB+23}, where each segment connects to at most two neighbors. Such designs limit scalability, as the length of the trap grows linearly with the number of qubits. Introducing junctions \cite{HensingerOSHYADMR06, BlakestadOVABLW09, WrightAFVDHPLDKSH13, ShuVBNVSB14} increases connectivity, allowing a segment to connect to more than two neighbors. Quantinuum has demonstrated this idea in its Helios architecture, where a ring-shaped storage region is connected to two linear legs via a single $\mathsf{X}$-shaped junction \cite{Ransford+26}. Increasing the number of junctions would allow more complex architectures, such as a two-dimensional quantum charge-coupled device (QCCD) \cite{LekitschWFMDWH17}.

\emph{Compilation} is the translation of a quantum circuit, which comprises abstract qubits and idealized gates, into a sequence of instructions executable on a specific trapped-ion device. This process begins by decomposing gates into the device's native gate set \cite{KreppelMOWHPSB23}, then mapping qubits to trapped ions, and finally determining a \emph{shuttling schedule}, the sequence of shuttling operations that brings the qubits of each gate into the gate segment. The schedule must respect the connectivity of the trap and the restrictions on where each operation may be applied. It is built from four primitive shuttling operations: \emph{Translate} moves the qubits of a segment to an empty neighbor, \emph{Separate} and \emph{Merge} split and rejoin them across two segments, and \emph{Swap} reverses the order of the qubits inside a single segment. \emph{Separate}, \emph{Merge}, and \emph{Swap} require precise calibration of the trap potentials and are therefore available only at the gate segment \cite{HilderPOSOLRMSP22}. This restriction holds throughout our datasets and evaluation, and it leaves the compiler more to plan, since every reordering of qubits has to be routed through the gate segment.

Dedicated compilers have been developed for linear segmented ion traps \cite{Wagner22, DurandauWMBSPB23, AshSakiTG22, WuW26} and for branched one-dimensional architectures with junctions \cite{KreppelMWHPSB24}. Recent algorithmic work carries this further, to other hardware designs \cite{SchoenbergerHBW24, SchoenbergerHBW25, SchoenbergerW25, SchoenbergerHSW25} and toward greater architectural generality \cite{BachSY26, RussonBYS26, ZhuWWW25}. In all of these compilers the routing strategy is designed by hand. That strategy can also be learned with reinforcement learning \cite{SchierRSSBHR26}, over a state and action space defined for the chip, with the agent retrained for each new chip.

We take a different route and employ LLMs. LLMs have been applied successfully across a wide range of problems \cite{ZhaoZLT+26}, and we also investigate whether they can be used for shuttling compilation. Two properties make them well suited to our problem. They perform new tasks from textual instructions or a few demonstrations \cite{Brown+20}, so the schedules of an existing compiler can serve directly as the examples to learn from. Furthermore, the trap is described to them in text, so moving to another trap changes only that description and not the structure of the LLM. In contrast, a network that encodes the trap in its structure has to be rebuilt for each new trap. 

We fine-tune five open-weight LLMs from four families (Llama 3.2 \cite{MetaAI24}, Qwen 3 \cite{QwenTeam25}, DeepSeek LLM \cite{Deepseek-AI24}, and Gemma 3 \cite{GemmaTeam25}) on schedules produced by the classical compilers for the linear \cite{Wagner22} and the branched one-dimensional \cite{KreppelMWHPSB24} architecture. The trap layout, the operation semantics, the gates to be executed, and the current ion positions are supplied as text. Each LLM then emits the shuttling operations, one gate at a time. The surrounding pipeline checks each generated operation against the trap rules and removes redundant operation pairs from the schedule in a post-processing step, but makes no routing decisions of its own.

Our contribution is the following: the resulting compilers produce valid schedules on both training architectures for circuits of up to 16 qubits, most reliably at the lower qubit counts. In 12\,\% of the compilations yielding a valid schedule, the best of ten runs outperforms the classical baselines \cite{Wagner22, KreppelMWHPSB24} by up to 21\,\% after post-processing. Those baselines are the compilers that produced the training data. The number of runs that yield a valid schedule differs widely between the five LLMs and does not follow their size. One LLM completes a schedule in a single run on an unseen four-way branched layout, while no LLM completes one on two other unseen layouts. We report this as preliminary evidence of cross-architecture generalization. To our knowledge, this is the first use of LLMs to compile circuits into shuttling schedules for trapped-ion quantum computers, and the first comparison of different LLMs on the task.

The remainder of this paper is structured as follows. \autoref{sec:related} surveys prior work on shuttling compilation and on learning-based routing. \autoref{sec:compiler} presents our LLM-based approach. \autoref{sec:evaluation} reports its evaluation on benchmark circuits and on unseen trap layouts, and \autoref{sec:conclusion} summarizes the contributions and outlines future work.

\section{Related Work}
\label{sec:related}

Any qubit of a quantum circuit may in principle interact with any other. However, the connectivity of a real device is limited, so only part of the required interactions can be carried out directly. Circuits are therefore expressed and optimized in compilation stacks that support several architectures, among them Qiskit \cite{JavadiAbhariTKWLGMNBCJG24}, t$\vert$ket$\rangle$ \cite{SivarajahDCSED20}, and Cirq \cite{CirqDevelopers25}, as well as more general-purpose platforms such as PennyLane \cite{BergholmISG+22}. Their output is then passed to a compiler for the target device. On shuttling-based trapped-ion systems that compiler maps the qubits to ions and derives a shuttling schedule. Because the number of legal operation sequences grows rapidly with the circuit size, these compilers rely on heuristics.

Most shuttling compilers are written for one specific trap architecture \cite{PinoDFGMABFHMRN21, MosesB+23, Ransford+26, SchmaleTBPKDOWB22, Wagner22, DurandauWMBSPB23, AshSakiTG22, WuW26, KreppelMWHPSB24, DurandauBSPMB26, ChangJCHL25, WebberHWH20, DaiBR24}, and that architecture in turn governs how much shuttling a circuit requires \cite{MuraliDBM20}. The MQT IonShuttler \cite{SchoenbergerHBW24, SchoenbergerHBW25, SchoenbergerW25, SchoenbergerHSW25} is tied to one trap architecture, a two-dimensional QCCD grid. It solves small instances exactly through Boolean satisfiability \cite{SchoenbergerHBW24} and larger ones heuristically \cite{SchoenbergerHBW25, SchoenbergerW25, SchoenbergerHSW25}. Compilers have also been built for other trapped-ion architectures. For trapped-ion systems assembled from modules linked by entanglement, movement between modules dominates the cost \cite{WuZWW25}. On two-dimensional machines, TrapSIMD aggregates intra-trap shifts and junction crossings so that ions move in parallel \cite{RuanZFLCHHHHPD25}. Other work covers more than one layout. SHAW and SHAPER run a heuristic derived from SABRE \cite{LiDX19} on a position-graph abstraction of the hardware \cite{BachSY26}, and a later variant extends it to larger circuits \cite{RussonBYS26}. S-SYNC treats the whole device as one fixed coupling graph and turns shuttling into the insertion of SWAP gates \cite{ZhuWWW25}. Compilers can also be generated automatically: a description of the device instantiates a parametric search algorithm \cite{MolaviXCTA26}. In all of these compilers the routing strategy is designed by hand.

The routing logic has also been replaced by a learned model. A reinforcement-learning agent interacts with a simulator of the device and learns to schedule ion movements between registers and through junctions on QCCD chips \cite{SchierRSSBHR26}. Its action set does not include the separation, merge, and swap operations of the segmented ion traps considered here.

Routing has also been learned on platforms other than ion traps. On a fixed coupling graph the qubits keep their positions and are brought together by inserting SWAP gates, and learning has been applied to that insertion and to the initial mapping \cite{PalerSFA23, PozziHSM22, PascoalFA24, TangDKFKS24, SinhaAS22, FanGL22, ZhouFL22}. On processors assembled from several chips or cores, agents assign the qubits to units \cite{RussoPPAC25, SangHH26, ZeynaliB25}, and on processors linked by entanglement another targets the reduction of execution time \cite{PromponasMCPST25}. Neutral atom arrays move their qubits as well, and there an agent plans the arrangement of the atoms before each layer of gates \cite{NakajiWHCPKA25}. The elementary operations in these settings differ from the ones considered in this paper.

All learned compilers mentioned above share the same formulation: an agent chooses one operation at a time from an action space fixed for the device, guided by a reward or a tree search. Our formulation differs in four respects. First, the unit of prediction is a gate rather than an operation: the LLM receives a trap state and emits the whole sequence that carries the circuit to the next gate execution. Second, the trap graph and the operation set are read from the instruction at inference rather than built into the policy, and the action space changes with the trap state. Third, the legal operations are listed in the instruction rather than enforced by masking, and a response containing an illegal one is discarded in full before the instruction is resubmitted. Fourth, the training signal is neither a reward nor a search, but the schedules a classical compiler has already produced.

\section{Compiler}
\label{sec:compiler}
\begin{figure}[!t]
	\centering
	\includegraphics[width=\columnwidth, keepaspectratio]{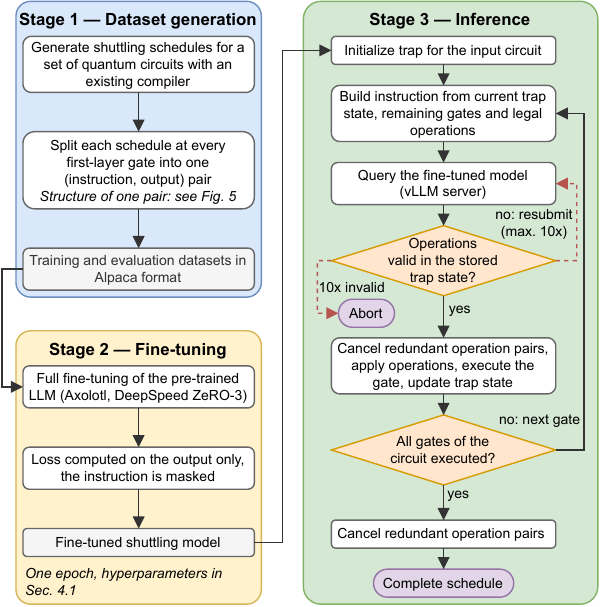}
	\caption{Algorithmic overview of the three phases of our compiler: dataset generation, fine-tuning, and inference. Inference runs two nested loops, an inner one that resubmits an instruction whenever the generated operations fail validation and an outer one that advances to the next gate. The structure of an instruction and the response to it is shown in \autoref{fig:example}.}
	\label{fig:flowchart}
\end{figure}

This section explains how we fine-tune LLMs so that they generate a shuttling schedule for a given circuit and trap. The process comprises three phases, dataset generation, fine-tuning, and inference, and is summarized in \autoref{fig:flowchart}. A fourth part reports the alternative designs we explored and the constraints they revealed.

\subsection{Dataset Generation}
\label{subsec:dataset-generation}
\begin{figure}[!t]
	\centering
	\includegraphics[width=\columnwidth, keepaspectratio]{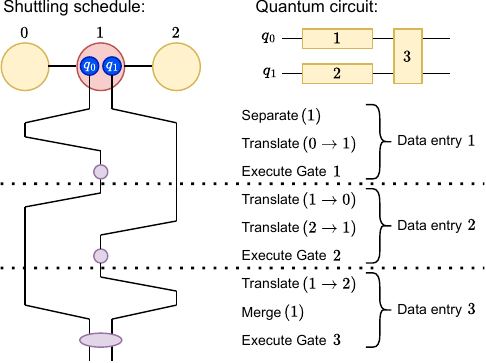}
	\caption{Shuttling schedule of a quantum circuit with two qubits and three gates on a linear architecture consisting of three segments. Both qubits are initially located in gate segment 1. The solid lines indicate the qubit movements between the segments, the purple dots denote gate executions, and the operations performed are annotated alongside the lines. For dataset generation, the schedule is split at the dotted lines into data entries, each holding the operations between two consecutive gates.}
	\label{fig:data-entry}
\end{figure}

The examples needed for fine-tuning are shuttling schedules, one of which is shown in \autoref{fig:data-entry}. Since the physical movements depend only on which qubits a gate acts on, and not on what that gate does, its function is omitted from this representation. Schedules can be generated in different ways, for example by a search that a reward function guides toward valid solutions, or by employing an existing compiler, which is the approach taken in this paper. Larger and more complex traps, as well as higher qubit counts, demand more shuttling operations per gate. This makes the generation of valid shuttling schedules harder for the LLM to learn and increases the amount of data required for fine-tuning.

\begin{figure}[!t]
    \centering
    \begin{subfigure}{1.0\columnwidth}
		\centering
		\includegraphics[width=\linewidth, keepaspectratio]{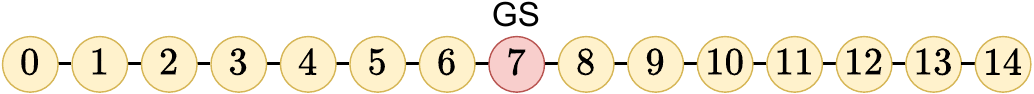}
		\caption{Linear architecture with 15 segments.}
		\label{fig:linear-trap}
	\end{subfigure}\\[6pt]
	\begin{subfigure}{1.0\columnwidth}
		\centering
		\includegraphics[width=\linewidth, keepaspectratio]{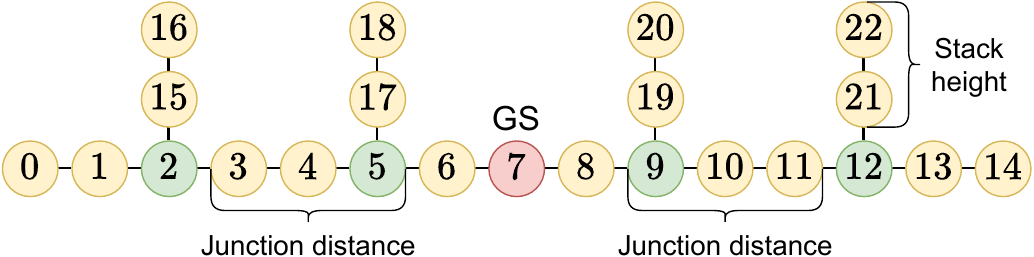}
		\caption{Branched one-dimensional architecture with 23 segments. The junction distance is $d = 3$, so consecutive junctions on the same side of the gate segment are three segments apart, and the stack height is $s = 2$, so each junction carries a stack of two qubit segments.}
		\label{fig:junction-trap}
	\end{subfigure}
    \caption{Graph representations of ion-trap architectures. Red vertices are gate segments (GS), yellow vertices qubit segments, and green vertices junctions, which are the only vertices connected to more than two others. The numbers are unique vertex identifiers.}
    \label{fig:traps}
\end{figure}

The trap is modeled as a graph $G = (\mathcal{V}, \mathcal{E})$, whose vertices $\mathcal{V}$ are the trap segments and whose edges $\mathcal{E}$ are the physical connections between neighboring segments. Each vertex is a gate segment, a qubit segment, or a junction. We focus on the linear and branched one-dimensional architectures, for which example graphs are shown in \autoref{fig:traps}.

From the generated shuttling schedules, we construct two datasets: one for training and one for evaluation. The evaluation dataset is smaller and is used during training to assess model performance on unseen data. Both hold schedules from different architectures, so that a single LLM is fine-tuned across them. We cut the schedules at the gate executions, as illustrated in \autoref{fig:data-entry}, so that each data entry holds the operations between two consecutive gates. The first entry starts from the initial trap state and the full circuit and ends at the first gate execution. Every later entry starts from the trap state that this execution produced, with the executed gate removed from the circuit, until the whole schedule has been consumed. Each data entry follows the Alpaca format \cite{TaoriGZDLGLH23}, consisting of an instruction and an output.

\subsubsection{Instruction}
\label{subsubsec:instruction}
The instruction encodes everything the LLM needs in order to reach the next gate execution. It describes the trap graph, the gates of the circuit, the available shuttling operations together with the constraints on them, the goal, and the current situation. Since LLMs operate on text, all of this is written out as a prompt.

The graph is given as its sets of vertices and edges, each vertex by a unique natural number and each edge by its two endpoints. Vertices along a linear section are numbered consecutively, so that $v-1$ and $v+1$ are the neighbors of $v$ within that section. A segment has a maximum capacity, and the qubits it holds form a linear chain. Every qubit therefore has a position in that chain, counted from zero, and is referenced as a tuple $[v, p]$ of its vertex number and position.

\begin{figure}[!t]
    \centering
    \includegraphics[width=\columnwidth, keepaspectratio]{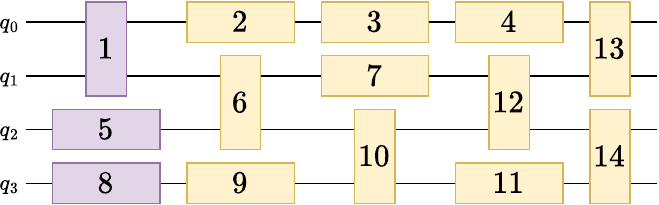}
    \caption{Quantum circuit with 4 qubits and 14 gates. Purple gates form the first layer, whereas yellow gates belong to deeper layers. The numbers give the order in which the gates on each qubit are executed.}
	\label{fig:quantum-circuit}
\end{figure}

In addition, the instruction specifies the gates of the circuit to be executed on the given architecture. Each gate targets specific qubits, and is executable only if all of them reside in the same gate segment and that segment holds no others. As illustrated in \autoref{fig:quantum-circuit}, the circuit is partitioned into layers. The first layer holds the gates whose qubits have no predecessor, whereas the gates in deeper layers become executable only once their predecessors have run. Each gate carries a number, assigned per qubit in ascending order. Gates on the same qubit must follow that order, while gates on disjoint qubits may be executed in any order. Finally, replacing the qubits named in the gate descriptions by the tuples $[v, p]$ of their positions links the circuit to the trap.

To modify the trap state, the LLM requires a description of the operations available to it: \emph{Translate}, \emph{Separate}, \emph{Merge}, and \emph{Swap}, together with gate execution. Gates can be executed only on the gate segments, whereas separation, merge, and swap may, depending on the hardware, also be permitted on qubit segments or on a subset of them.

In segmented trapped-ion architectures, junctions are formed by the intersections of electrodes, enabling qubits to be routed between different linear sections. A junction cannot hold a qubit for any length of time \cite{BlakestadODWBLW11}, so qubits must pass through it without stopping. It is therefore occupied only while qubits move from one translation to the next, and stands empty during every other operation. For \emph{Separate} and \emph{Merge}, the adjacent vertices involved must therefore not be junctions either. Reversing direction inside a junction is infeasible as well, since the electric fields are engineered for unidirectional traversal rather than oscillatory motion \cite{WrightAFVDHPLDKSH13, MokhberiSW17}. The operations are formally defined as follows:

\begin{itemize}
    \item \textbf{Translate:} Let $v \in \mathcal{V}$ hold qubits and let $u \in \mathcal{V}$ be an empty neighbor, so that $\{v, u\} \in \mathcal{E}$. All qubits of $v$ are moved to $u$. After translation, $v$ is empty, and if $u$ is a junction, the next translation out of $u$ must go to a neighbor $w \neq v$.
    \item \textbf{Separate:} Let $v \in \mathcal{V}$ be a vertex at which \emph{Separate} is permitted, containing $n \geq 2$ qubits, with empty neighbors $v-1$ and $v+1$. The qubits at positions $0, \dots, \lceil \tfrac{n}{2} \rceil - 1$ in the qubit chain of $v$ are moved to $v-1$, while the remaining qubits are moved to $v+1$. After separation, $v$ is empty.
    \item \textbf{Merge:} Let $v \in \mathcal{V}$ be an empty vertex at which \emph{Merge} is permitted, with neighbors $v-1$ and $v+1$ containing $m$ and $n$ qubits, respectively, such that $m+n$ does not exceed the maximum segment capacity. The qubits from $v-1$ and $v+1$ are moved into $v$, where positions $0, \dots, m-1$ are occupied by the qubits from $v-1$ and positions $m, \dots, m+n-1$ by the qubits from $v+1$. After the merge, $v-1$ and $v+1$ are empty.
    \item \textbf{Swap:} Let $v \in \mathcal{V}$ be a vertex at which \emph{Swap} is permitted, containing at least two qubits. The order of the qubit chain in $v$ is reversed.
    \item \textbf{Execute Gate:} The gate is performed in the gate segment, once the conditions stated above are satisfied.
\end{itemize}

The instruction then states the goal, to execute all gates with the fewest operations, and the current situation, given as four lists:

\begin{itemize}
    \item \textbf{Qubit positions:} The qubits in the trap, each given as its tuple $[v, p]$.
    \item \textbf{First-layer gates:} The gates of the first layer, with the tuples of the qubits they act on.
    \item \textbf{Deeper-layer gates:} The gates that become executable next, again with their qubit tuples. To keep the focus on near-term execution, only the next executable gates are listed, rather than all remaining gates of the circuit.
    \item \textbf{Allowed shuttling operations:} The operations that are legal in the current trap state, so that the LLM can select a valid first step without ambiguity.
\end{itemize}

The instruction closes with the task itself: produce a sequence of operations that lets one of the first-layer gates be executed. After each operation, the LLM should also report the updated qubit positions, the first-layer gates with their updated tuples, and the operations now allowed. Requiring these outputs at each step forces the LLM to record where the qubits now are and what the operation it has just chosen changed. The deeper-layer gates are not repeated, since none of them can run until such a gate has been executed.

\subsubsection{Output}
\label{subsubsec:output}
The output is the part of the schedule that carries the trap from the given state to the execution of one first-layer gate. It lists the operations one after another, each followed by the updated qubit positions, the first-layer gates, and the operations now allowed, so that the effect of every operation is visible. The closing \emph{Execute Gate} needs no such update and ends the output. If the gate is already executable, the output consists of that single operation. Otherwise several shuttling operations precede it.

\subsection{Fine-tuning with Axolotl}
\label{subsec:fine-tuning}
We fine-tune pre-trained LLMs from Hugging Face \cite{Wolf+20} with \emph{Axolotl} \cite{Axolotl23}. It supports full fine-tuning, Low-Rank Adaptation (LoRA) \cite{HuSWALWWC22}, and Quantized LoRA (QLoRA) \cite{DettmersPHZ23}. We use full fine-tuning, since with LoRA and QLoRA the output departed from the expected format, as \autoref{subsec:alternative_approaches} discusses. \emph{Axolotl} also gives access to the performance optimizations described below, without which such a run would not fit on the GPUs available to us.

\subsubsection{Performance Optimizations}
\label{subsubsec:performance}
Fine-tuning at this scale requires reducing memory usage and parallelizing execution. We do both with \emph{DeepSpeed} \cite{RasleyRRH20}, employing the Zero Redundancy Optimizer (ZeRO) at optimization level 3 \cite{RajbhandariRRH20}, which partitions the optimizer states, the gradients, and the model parameters across the GPUs. \emph{DeepSpeed} gathers the parameters again when a forward or backward pass needs them. The memory held on the GPU is kept small: the parameters are held in \texttt{bfloat16}, and \emph{gradient checkpointing} \cite{ChenXZG16} discards the activations after the forward pass and recomputes them when the backward pass needs them. We additionally apply ZeRO Offloading \cite{RenRARYZLH21}, which moves the optimizer states and the parameters that are not currently in use to the host CPU and its RAM.

As described above, the entries differ widely in length, from a single operation to long shuttling sequences. Since fine-tuning requires sequences of uniform length, shorter ones are padded with special tokens. Rather than pad every entry to the longest, we apply \emph{multipacking} \cite{KunduLWGM24}, which trains on several short entries at once, up to the model’s context window. Attention is computed with \emph{FlashAttention 2} \cite{Dao24}, a memory-efficient and fast implementation that raises the length a single GPU can hold. Entries that still exceed the length a single GPU can hold are split across several GPUs by \emph{sequence parallelization} \cite{KorthikantiCLMASC23}, which uses blockwise ring attention \cite{LiuZA24} to overlap communication with computation.

Fine-tuning itself proceeds in two steps, pre-processing and training. \emph{Axolotl} runs both from a single YAML file, which names the optimization settings and the paths to the LLM, the two datasets, and a separate \emph{DeepSpeed} configuration.

\subsubsection{Pre-processing}
\label{subsubsec:pre-processing}
\emph{Axolotl} first validates the YAML file, checking that the required libraries are present and that the selected optimizations are compatible. It then initializes the tokenizer and formats both datasets. The instruction states the problem while the output gives the solution, so the instruction tokens are masked and the loss is computed on the output alone. This pushes the LLM to solve the task rather than reproduce the prompt. For \emph{multipacking}, entries are concatenated up to a maximum length, each keeping its own loss mask. The processed dataset, including tokenized inputs, loss masks, and associated metadata, is saved in Apache Arrow format for use during training.

\subsubsection{Training}
\label{subsubsec:training}
\emph{Axolotl} loads the LLM, tokenizer, and pre-processed datasets using Hugging Face's \texttt{datasets} library and distributes them according to the selected optimizations. It then configures the \texttt{Trainer} from Hugging Face's \texttt{transformers} library and connects it to \emph{DeepSpeed}. While \emph{DeepSpeed} handles memory optimizations and parallelization, the \texttt{Trainer} manages the training loop, which loads micro-batches of tokenized data. To reduce memory usage further, each iteration uses a batch size of one. For each entry of the training dataset, the forward pass computes the loss on the output, and backpropagating that loss yields the gradients. We use gradient accumulation over eight batches to mitigate gradient noise and to improve convergence despite the small batch size. The accumulated gradients are applied by the AdamW optimizer \cite{LoshchilovH19} to update the network weights. Alongside the training loop, the \texttt{Trainer} logs progress, saves checkpoints, and measures the loss on the evaluation dataset at regular intervals, which tracks the performance on unseen data. We fine-tune for a single epoch, which lets us spend the same budget on a larger dataset and so on a greater variety of examples. Upon completion, the fine-tuned LLM is saved for inference.

\subsection{Inference with vLLM}
\label{subsec:inference}

\begin{figure*}[!t]
	\centering
	\begin{subfigure}{1.0\textwidth}
		\centering
		\includegraphics[width=\linewidth, keepaspectratio]{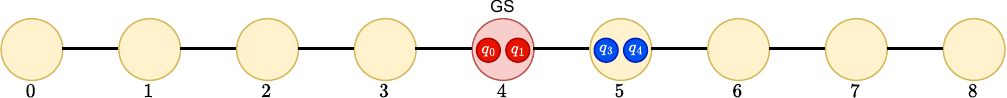}
		\caption{Initial trap state.}
		\label{fig:example-state}
	\end{subfigure}\\[6pt]
	\begin{subfigure}{1.0\textwidth}
		\centering
		\includegraphics[width=\linewidth, keepaspectratio]{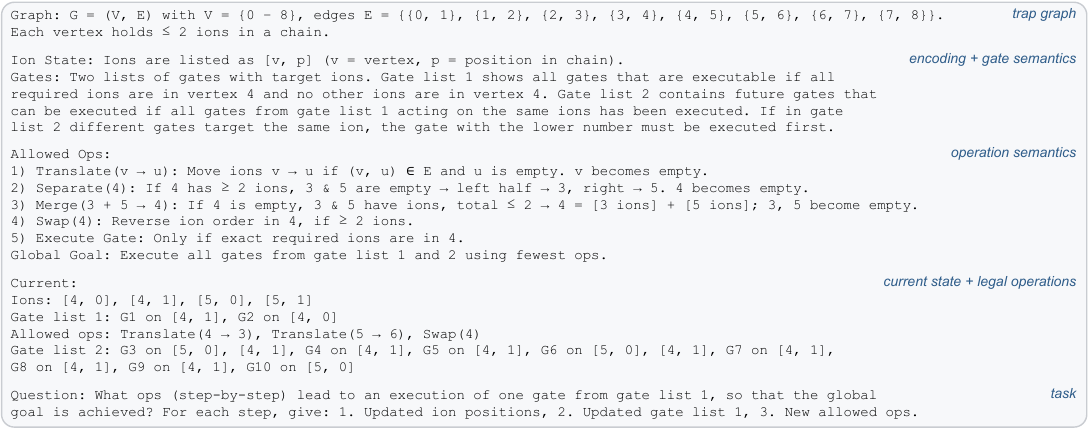}
		\caption{Instruction submitted to the \emph{vLLM} server, reproduced verbatim. It serializes the trap graph, the operation semantics, the current ion positions, the first-layer and deeper-layer gates, and the operations that are legal in this state. Only the \texttt{Current:} block changes between consecutive steps.}
		\label{fig:example-prompt}
	\end{subfigure}\\[6pt]
	\begin{subfigure}{1.0\textwidth}
		\centering
		\includegraphics[width=\linewidth, keepaspectratio]{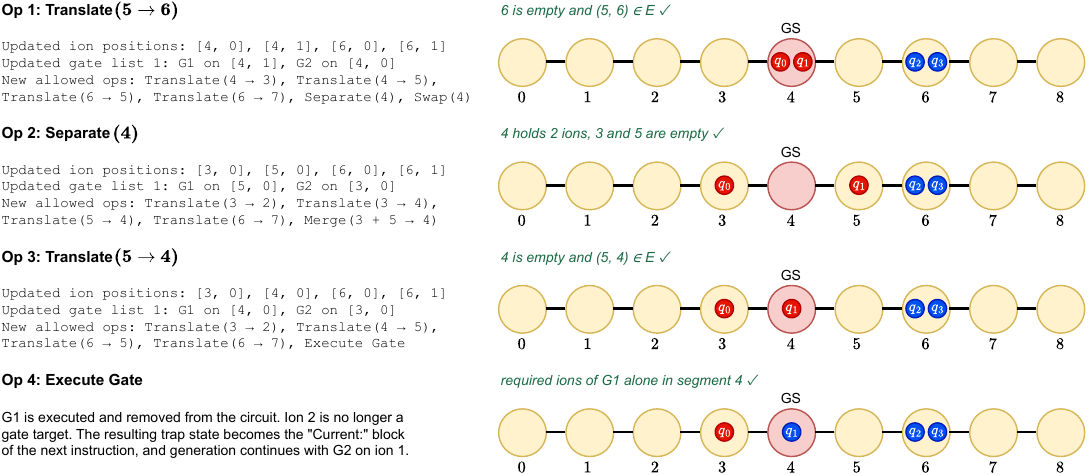}
		\caption{Response returned by the LLM, with the structure defined in \autoref{subsubsec:output}. Each generated operation is accompanied by the updated ion positions, the updated first-layer gate list, and the new set of legal operations. The graphs on the right give the resulting trap state. The annotations name the condition that validation checks before an operation is accepted, as in \autoref{fig:validation-reject}. After \texttt{Execute Gate}, gate $G_1$ is removed from the circuit and the resulting state becomes the \texttt{Current:} block of the next instruction.}
		\label{fig:example-response}
	\end{subfigure}
	\caption{Worked example of one inference step on a linear architecture with nine segments and four ions. Nodes are trap segments, edges give their connectivity, and ions sit inside the nodes. Ions drawn in red are the targets of the first-layer gates $G_1$ and $G_2$, while ions drawn in blue are stored. GS marks the gate segment, which is the only segment in which gates, separations, merges, and swaps can be executed.}
	\label{fig:example}
\end{figure*}

\begin{figure}[!t]
	\centering
	\begin{subfigure}{1.0\columnwidth}
		\centering
		\includegraphics[width=\linewidth, keepaspectratio]{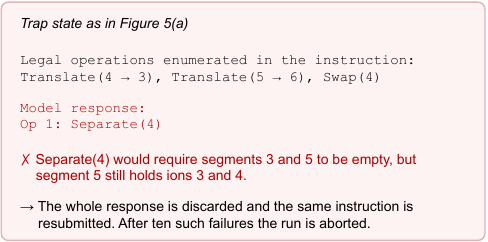}
		\caption{Validation. In the trap state of \autoref{fig:example-state} the instruction enumerates three legal operations. A response beginning with \texttt{Separate(4)} is rejected, because separation requires both neighbors of the gate segment to be empty while segment 5 is still occupied. Rejection discards the entire response, not the offending operation, and the same instruction is resubmitted.}
		\label{fig:validation-reject}
	\end{subfigure}\\[6pt]
	\begin{subfigure}{1.0\columnwidth}
		\centering
		\includegraphics[width=\linewidth, keepaspectratio]{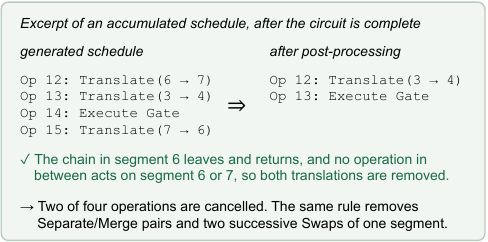}
		\caption{Post-processing. The schedule is scanned for operation pairs that cancel. Here a qubit is translated out of segment 6 and later translated back, with no intervening operation acting on segments 6 or 7, so both translations are removed.}
		\label{fig:validation-postproc}
	\end{subfigure}
	\caption{The two checks applied to the generated operations: validation of each response, and cancellation of redundant operation pairs in the schedule.}
	\label{fig:validation}
\end{figure}

We run the fine-tuned LLM with \emph{vLLM} \cite{KwonLZSZYGZS23}, which provides efficient multi-GPU support. Its \emph{PagedAttention} mechanism stores the attention keys and values in pages, keeping their memory usage low. A local \emph{vLLM} server loads the fine-tuned LLM and its tokenizer. The server exposes an OpenAI-compatible API accepting JSON-formatted HTTP requests. Each request contains an instruction in the same format as the training data, together with the maximum number of tokens to generate. Since an output may produce many operations, this limit is set as large as the LLM permits, at the values given in \autoref{subsec:setup}. The server returns the response in JSON format, from which we extract the generated output. Control tags such as \texttt{<think>} are stripped before the operations are read off.

Each response ends at the next first-layer gate, so a full schedule takes many requests, one of which is shown in \autoref{fig:example}. A Python script issues these requests and checks each output, since the LLM may emit operations that the architecture or the current state does not allow. The script tracks both the qubit positions in the trap and the gates of the circuit that remain. At initialization, it loads the quantum circuit, takes the initial qubit placement from the compiler that generated the training schedules, and determines the lists of first-layer gates, deeper-layer gates, and allowed shuttling operations. From this it builds the instruction and submits it to the server.

The returned output is valid if every operation obeys the constraints of \autoref{subsec:dataset-generation} and at least one first-layer gate has been executed. An invalid output, as \autoref{fig:validation-reject} illustrates, is discarded and the same instruction is resubmitted, so a retry draws another sample from the LLM rather than repairing the rejected output. Each retry is an independent generation with the same success probability, so the chance that at least one of them succeeds grows with their number, while the set of reachable responses is unchanged. After ten failed retries on the same gate, the run is abandoned. In a post-processing step, the script removes redundant operation pairs from a valid output, as in \autoref{fig:validation-postproc}: a qubit translated back and forth between two segments, a merge and a separation on the same qubits in succession, or two successive swaps. The script then updates the qubit positions, drops the executed gate from the circuit, and builds the next instruction from the new state. This repeats until all gates have been executed. The same post-processing step then runs once more over the finished schedule, where it can also remove operation pairs that span two outputs.

The server could also return several outputs per instruction, from which the shortest valid one would be kept. We generate a single output instead, because one already takes 24 seconds at the median and up to two and a half minutes, depending on the LLM and on the number of operations the gate requires.

\subsection{Alternative Approaches and Design Insights}
\label{subsec:alternative_approaches}
Before settling on the pipeline described above, we ran two preliminary checks to investigate whether a learned policy is needed at all and whether it has to be fine-tuned.

We first tried to generate schedules without a learned policy, by sampling uniformly from the legal operations at each step. Every operation was valid by construction, so no validation or retry was involved, but the sequences rarely reached a gate at all. Most of the sampled operations brought the next gate no closer to execution. Qubits were translated that were already in place or that the next gates did not involve, the same qubit was moved back and forth between two segments, and chains were separated, merged, and swapped where none of this was needed. The reason is that legality says nothing about progress. Every legal operation is available whether or not it advances the circuit execution, and one that undoes earlier progress is as legal as one that makes it. Respecting the constraints is therefore not enough, and the length of the schedule depends on which of the permitted operations is chosen.

The second check asked whether fine-tuning is needed at all. We gave the LLMs in their released state an instruction of the kind shown in \autoref{subsubsec:instruction}. Their responses were dominated by reasoning and commentary, and we extracted the operations from the surrounding text by hand. Most of them violated the rules of the trap, moving qubits into segments that had not been cleared or translating along edges absent from the trap graph. Fine-tuning is therefore necessary, and the released LLMs were left at this check.

Three alternative designs were explored in addition, each revealing a constraint that shaped the pipeline. First, we gave each shuttling operation its own data entry, so that an output holds a single operation rather than the whole sequence up to the next gate. The entries then became shorter, but the LLMs picked valid operations almost at random instead of ones that advance the circuit. The LLM evidently needs the whole sequence in one entry in order to learn to make progress. Second, we replaced the layer-aware encoding of the instruction with the full remaining circuit in OpenQASM \cite{CrossBSG17, CrossJADBHRSSGJ22} format. The state descriptions then grew much longer, and the LLM began to target arbitrary deeper-layer gates. A compact, layer-aware encoding is therefore preferable. Third, we fine-tuned with LoRA and QLoRA rather than adapting all weights. The LLMs then produced step-by-step reasoning that departed from the expected format and did not always respect the shuttling rules, which is why the final pipeline uses full fine-tuning. Since the capacity of an adapter grows with its rank and with the set of adapted weight matrices, a larger adapter might well close this gap, leaving low-rank adaptation a resource-efficient direction for future work.

\section{Evaluation}
\label{sec:evaluation}
\begin{figure}[!t]
	\centering
	\begin{subfigure}{0.49\columnwidth}
		\centering
		\includegraphics[width=\columnwidth, keepaspectratio]{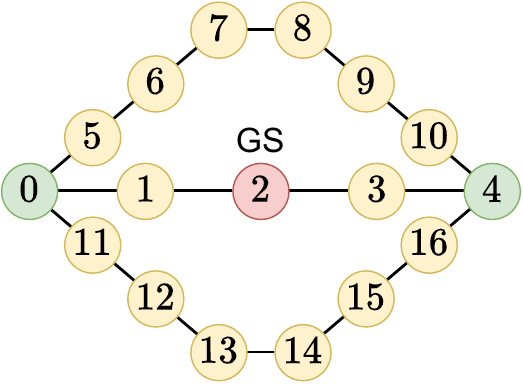}
		\caption{Theta architecture with three-way junctions.}
        \vspace{2mm}
		\label{fig:theta-trap}
	\end{subfigure}
	\hfill
	\begin{subfigure}{0.49\columnwidth}
		\centering
		\includegraphics[width=\columnwidth, keepaspectratio]{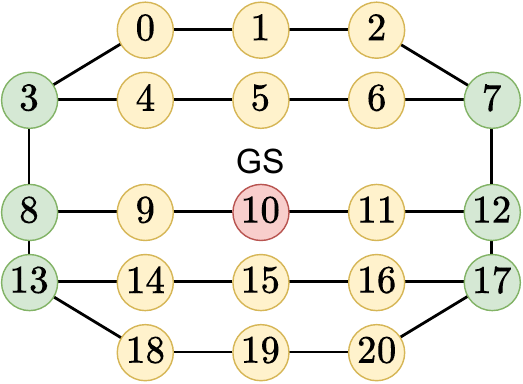}
		\caption{Multi-linear architecture with three-way junctions.}
        \vspace{2mm}
		\label{fig:multi-linear-trap}
	\end{subfigure}
	\begin{subfigure}{\columnwidth}
		\centering
		\includegraphics[width=\columnwidth, keepaspectratio]{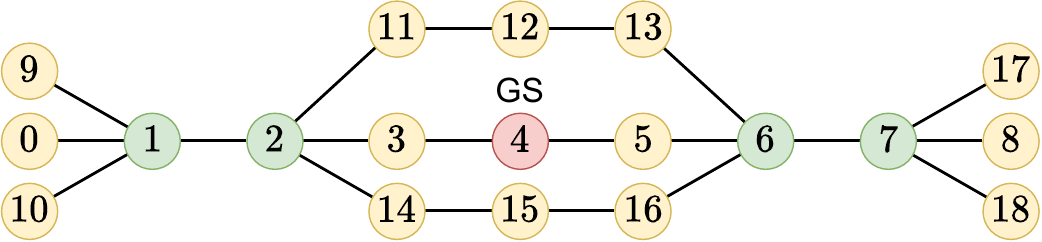}
		\caption{Four-way architecture, with junctions of degree four.}
		\label{fig:four-way-junction-trap}
	\end{subfigure}
	\caption{Graph representations of the three unseen layouts on which generalization is evaluated. Red vertices are gate segments (GS), yellow vertices qubit segments, and green vertices junctions, and the numbers are unique vertex identifiers. In each of them the number of qubit segments was scaled to the qubit count of the test circuit.}
	\label{fig:evaluation-traps}
\end{figure}

This section examines whether the fine-tuned LLMs produce valid shuttling schedules, how their schedules compare with those of the classical compilers they learned from, and how far their ability extends beyond the architectures they were trained on. The benchmark circuits are compiled for the two architectures seen during training and for three unseen layouts, shown in \autoref{fig:evaluation-traps}. The following subsections describe the experimental setup, present the results, and discuss the limitations.

\subsection{Setup}
\label{subsec:setup}
The three phases of \autoref{sec:compiler} are described in turn, followed by a discussion of the comparability of our results to other compilers.

\subsubsection{Dataset generation}
\label{subsubsec:setup-datasets}
We generated one training and one evaluation dataset, each covering the linear and the branched one-dimensional architectures. Both architectures had a single gate segment, and the \emph{separate}, \emph{merge}, and \emph{swap} operations were permitted only there. Every gate therefore requires its qubits to be moved into that single segment, so the LLM has more operations to plan for it. Each trap graph carried as many qubit segments on each side of the gate segment as the circuit had qubits. For the branched architecture we additionally varied two parameters of the trap, both shown in \autoref{fig:junction-trap}. The \emph{stack height} $s$ is the number of qubit segments per stack, and the \emph{junction distance} $d$ is the spacing between consecutive junctions on the same side of the gate segment. A branched architecture design is written as the pair $(s,d)$.

The circuits ranged from 2 to 16 qubits on the linear architecture and from 3 to 16 on the branched one. We restricted the datasets to these sizes because larger circuits need longer shuttling sequences, and so more VRAM and more GPUs than were available to us. The underlying quantum circuits were randomly generated with Qiskit \cite{JavadiAbhariTKWLGMNBCJG24} from single- and two-qubit gates, each circuit of depth five, and were then compiled into a trapped-ion-compatible format \cite{KreppelMOWHPSB23}. Each data entry shows only the trap state, the first-layer gates, the gates that become executable next, and the shuttling operations allowed in that state. A deep circuit therefore yields entries of the same form and comparable content as several shallow ones, and a greater depth would not extend what an entry contains. Shuttling schedules for these circuits were produced by the compilers \cite{Wagner22} and \cite{KreppelMWHPSB24} for the linear and the branched architecture, respectively. These two compilers, both strongly optimized heuristics, also serve as the baseline throughout this paper. The branched compiler emits no swap at all, so the branched training schedules contain none either, whereas the linear ones do.

For the linear architecture, the training dataset holds 120 shuttling schedules per qubit count and the evaluation dataset 30 of them. For the branched architecture the counts are eight and two per architecture design and qubit count. Since a schedule is cut into one entry per gate, both datasets hold comparable numbers of entries from the two architectures. \emph{Axolotl} shuffles both before fine-tuning, so the architectures are interleaved rather than presented in blocks. Training and evaluation dataset share no circuit, and the evaluation dataset is used during fine-tuning to monitor performance on unseen data.

\subsubsection{Fine-tuning}
\label{subsubsec:setup-fine-tuning}
Fine-tuning ran on a single node with eight H100 GPUs, each providing 80\,GB of VRAM. We fine-tuned five open-weight LLMs from four widely used families, spanning roughly an order of magnitude in parameter count: Llama 3.2 with 3 billion parameters \cite{MetaAI24}, Qwen 3 with 4 and with 14 billion \cite{QwenTeam25}, DeepSeek LLM with 7 billion \cite{Deepseek-AI24}, and Gemma 3 with 27 billion \cite{GemmaTeam25}. DeepSeek has a native context window of only 4{,}096 tokens, and we extended that window to 40{,}000 tokens by full fine-tuning at that sequence length. Larger LLMs such as DeepSeek V3 \cite{Deepseek-AI24_2}, DeepSeek R1 \cite{GuoYZ+25}, or Llama 4 \cite{MetaAI25} exceed what fits into a full fine-tuning run on our eight H100 GPUs and are left for future work.

We also fine-tuned GPT-OSS with 20 billion parameters \cite{OpenAI25}, which required replacing DeepSpeed with FSDP2 \cite{Zhao+23}. Its responses consistently omitted the first one or two shuttling operations, so they could not be assembled into valid schedules. We therefore report no inference results for it.

\subsubsection{Inference}
\label{subsubsec:setup-inference}
Inference ran on a single A100 GPU per LLM, on the Mogon clusters of Johannes Gutenberg University. Llama, Qwen, and DeepSeek used the 40\,GB cards of Mogon NHR, and the larger Gemma the 80\,GB cards of Mogon KI. The two baseline compilers have been evaluated on a circuit library of 153 instances \cite{Zhou19}. In this paper, we selected three of them, all used in prior work \cite{KreppelMWHPSB24, KreppelMOWHPSB23, ZhouLF20, CowtanDDKSS19}: \textit{4mod5-bdd\_287} (7 qubits, 106 gates), \textit{mini\_alu\_305} (10 qubits, 263 gates), and \textit{cnt3-5\_179} (16 qubits, 230 gates), the gate counts being those after compilation \cite{KreppelMOWHPSB23}. We selected these circuits for two reasons. First, they span different qubit counts, so the results show how the approach behaves as the number of qubits grows. Second, within each qubit count we took the circuit with the lowest gate count. Every request to the \emph{vLLM} server returns the operations for a single gate and takes 24 seconds at the median, so a circuit of 263 gates already costs hours per run. The results below therefore characterize the approach at the lower end of the gate counts occurring at each of the three qubit counts.

For each circuit, we used the fine-tuned LLMs to generate shuttling schedules for the linear architecture and for 10 (\textit{4mod5-bdd\_287}), 18 (\textit{mini\_alu\_305}), and 29 (\textit{cnt3-5\_179}) branched architectures $(s,d)$. To assess generalization, we additionally evaluated the LLMs on the previously unseen layouts shown in \autoref{fig:evaluation-traps}. As in the training data, each graph contained a single gate segment, and the \emph{separate}, \emph{merge}, and \emph{swap} operations were executable only within that segment. Every run started from the initial qubit placement of the compiler that generated the training schedules, so where a baseline exists it differs from our LLMs only in the routing decisions that follow. The sampling temperature $T$ controls how random the generated output is, and we ran every LLM–circuit–architecture combination at $T=0.7$ and $T=1.0$ to cover a more and a less deterministic regime. We performed ten runs per combination and temperature. The maximum number of tokens generated per request was set to 40{,}000 for DeepSeek, 40{,}960 for the two Qwen variants, and 50{,}000 for Llama and Gemma.

\subsubsection{Comparability with other compilers}
\label{subsubsec:comparability}
A numerical comparison of operation counts is direct when the compilers share the same elementary operations and the same trap architecture. Our baselines \cite{Wagner22, KreppelMWHPSB24} meet both conditions, which is why we compare against them throughout. None of the shuttling compilers of \autoref{sec:related} meets both. Two of them are written for other trap architectures, and two others, whose abstractions cover ours, differ from us in the elementary operations.

The compiler trained with reinforcement learning \cite{SchierRSSBHR26} is written for two chips. One is plus-shaped, with four segment arms meeting at a single junction, and the other is Q-shaped, with two further segments attached by a junction to a ring of qubit segments. The MQT IonShuttler \cite{SchoenbergerHBW24, SchoenbergerHBW25, SchoenbergerW25, SchoenbergerHSW25} targets a two-dimensional QCCD grid. None of these is a linear or branched one-dimensional architecture, and neither compiler has been shown to compile for such a trap. Their operation sets follow those designs and contain no counterpart to our \emph{Separate}, \emph{Merge}, and \emph{Swap}. The exact Boolean-satisfiability formulation of the MQT IonShuttler \cite{SchoenbergerHBW24} is therefore no optimality reference for our schedules.

SHAW/SHAPER \cite{BachSY26} and S-SYNC \cite{ZhuWWW25} allow up to $n$ qubits in one segment, but translate differently. Where our \emph{Translate} moves a complete chain into an empty neighbor, theirs splits the outermost qubit from the chain and moves that qubit alone, joining it to the chain of the neighbor. The neighbor therefore need not be empty, only below its capacity, and translating a chain of $n$ qubits takes $n$ of their moves. They call that split a separation and that join a merge. Our \emph{Separate} and \emph{Merge} instead divide a chain across both neighbors and join two neighboring chains. With two qubits per segment, as in our evaluation, each of our \emph{Translate}, \emph{Separate}, and \emph{Merge} becomes up to two of their moves, one for each qubit of the chain, and \emph{Swap} coincides with their exchange of two neighboring qubits. This allows every schedule of ours to be expressed in their operations, while a single move of theirs requires several of ours.

An example, constructed by hand from the two operation sets, shows what the difference costs. Consider the path 0--1--2--3--4 with a capacity of two qubits per vertex and the gate segment at vertex 2, which holds the chain $[q_0, q_1]$ while vertex 3 holds $[q_2]$, and let a gate on $q_0$ be executed. With their operations, $q_1$ moves into the free place beside $q_2$, after which the gate can be executed, so one shuttling operation precedes it. No single operation of ours produces that state, and reaching the same trap state with our operations takes ten. A chain can only be divided once both neighbors of its segment are empty. So $q_2$ is first translated to vertex 4 in order to separate the chain in the gate segment. Then $q_0$ is translated to vertex 0, $q_1$ to vertex 1, and $q_2$ back to vertex 3, so that $q_1$ and $q_2$ can be merged. The merged chain is translated to vertex 3 and $q_0$ back into the gate segment. In contrast, our operations reach the same gate in three steps: vertex 3 is cleared to vertex 4, $q_0$ and $q_1$ are separated, and $q_0$ is translated back into the gate segment. Rewriting our own schedule with their moves raises it to four, since the separation alone becomes two of them. An operation count obtained this way depends on the conversion chosen to make the two counts comparable, and would describe the two operation sets rather than the two compilers. We therefore report no comparison against them.

\begin{figure*}[!t]
	\centering
	\includegraphics[width=\textwidth, keepaspectratio]{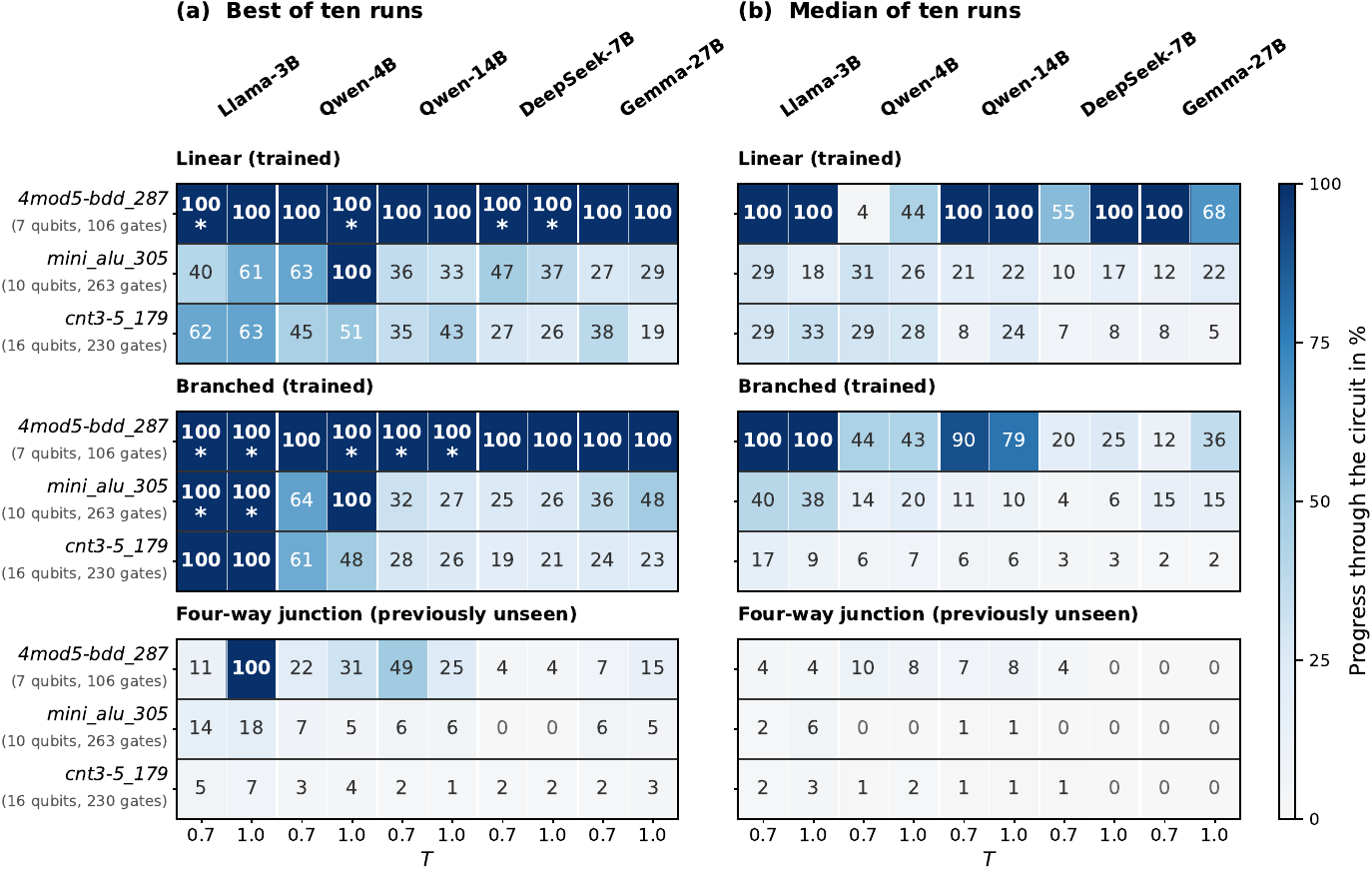}
	\caption{Progress of the LLM-generated shuttling schedules for every LLM, temperature, and circuit, on the two trained architectures and the four-way layout. Theta and multi-linear are omitted, since every one of their cells would be at or near zero. Each cell gives the percentage of the circuit's gates executed, in (a) by the most progressed of the ten runs and in (b) as the median over those ten. For the branched architecture each is taken over all tested $(s,d)$. The shading encodes that percentage on a scale shared by both panels, from $0\,\%$, no gate executed, to $100\,\%$, a complete and valid schedule. An asterisk in (a) marks cells holding at least one configuration whose best run needs fewer operations than its baseline \cite{Wagner22, KreppelMWHPSB24}. Eleven cells are starred, holding fourteen such configurations in total. Panel (b) carries no asterisk, since the median of the ten runs of a configuration is never a complete schedule shorter than the baseline.}
	\label{fig:results-heatmap}
\end{figure*}

\subsection{Results}
\label{subsec:results}
A run either produces a complete schedule for the whole circuit or stops after ten consecutive invalid outputs for the same instruction, in which case it has executed only part of the circuit. Only a completed run yields an operation count. A configuration is a fixed combination of LLM, temperature, circuit, and architecture. Because generation is non-deterministic, a single count quoted for a configuration is the best of ten runs, meaning the smallest count among them. The extra runs cost time but not schedule quality, and a shorter schedule directly improves execution fidelity on the device \cite{HilderPOSOLRMSP22}. By contrast, the baseline compilers are deterministic and need only one run per configuration.

All operation counts we report are taken after the post-processing step of \autoref{subsec:inference}. Following the convention of the baseline compilers \cite{Wagner22, KreppelMWHPSB24}, every count in this paper includes the gate executions alongside the shuttling operations, so that a schedule for a circuit with $g$ gates counts $g$ of them. On the branched architecture the baselines differ between designs $(s,d)$, so every comparison is made at the same $(s,d)$. We report the cost of a compilation both in generated tokens and in runtime. Token counts are hardware-independent and commonly used as a billing metric for LLMs \cite{VelascoTOG25}, whereas runtime depends on GPU type, the number of GPUs, and their interconnects, but shows how long a compilation takes.

Configurations that produced no complete schedule have no operation count, and are characterized by the proportion of the circuit's gates their runs executed. That proportion is given in \autoref{fig:results-heatmap}, for the most progressed and for the median run, on the two trained architectures and the four-way layout. Since the ten runs of a configuration can vary widely, \autoref{tab:linear} to \autoref{tab:branched-b} in \hyperref[sec:appendix]{Appendix~\ref*{sec:appendix}} report, for every configuration that produced a schedule, our best-of-ten and baseline counts together with the number of completed runs and the mean and standard deviation over those runs. They also list the raw counts before post-processing, the retries and response tokens required, and the runtime.

We first summarize what the compilers achieve, then report the per-circuit results on the trained architectures and which shuttling operations the schedules are made of. The results on the unseen layouts follow, then the effect of the sampling temperature, how far post-processing shortens the schedules, and what a compilation costs. At the end, we discuss how an LLM can produce schedules shorter than the baselines it learned from.

\begin{figure*}[!t]
	\centering
	\includegraphics[width=\textwidth, keepaspectratio]{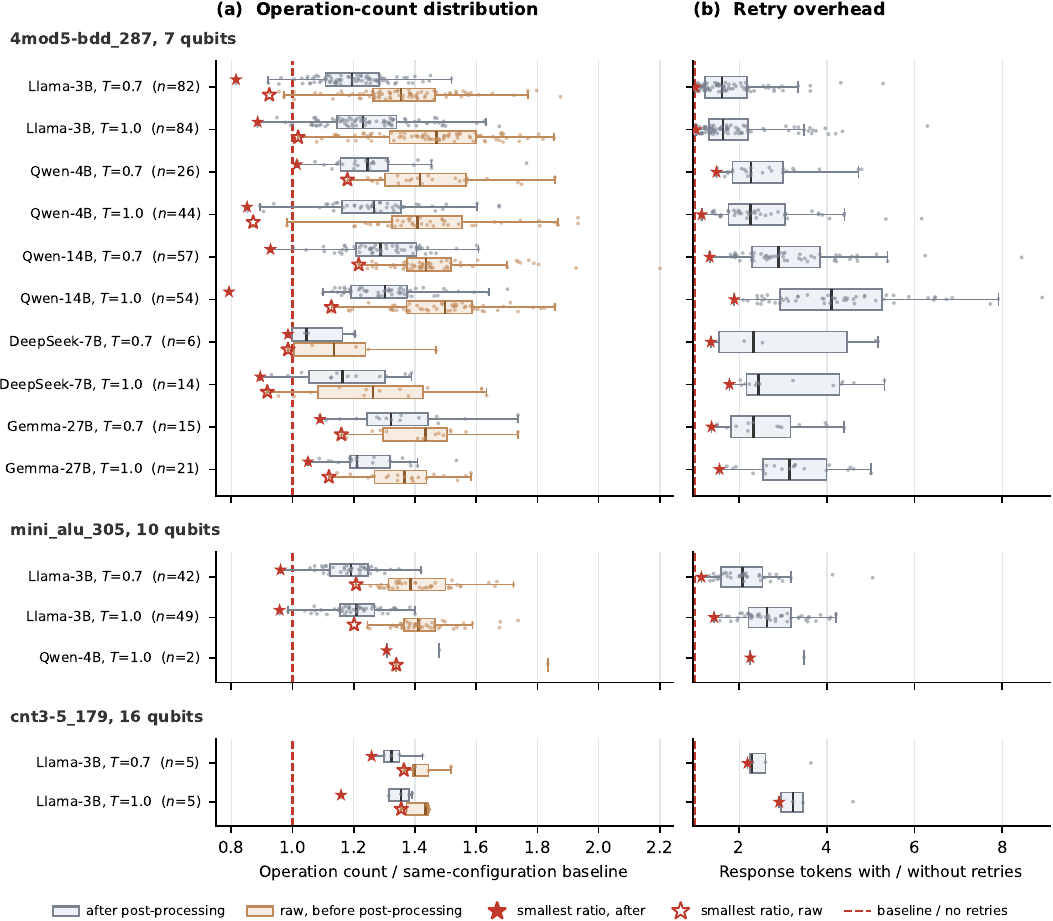}
	\caption{Distribution of the 506 runs that produced a complete schedule and for which a baseline is defined, split by benchmark circuit and grouped by LLM and sampling temperature, with $n$ the number of such runs. A group appears only if at least one of its runs completed, so the lower two blocks contain fewer rows. \textbf{(a)} Operation count of each run divided by that of the baseline at the same $(s,d)$, before and after the post-processing step. The dashed line marks parity. A filled star marks the smallest post-processed ratio in that group and an open star the smallest raw ratio, each taken over all configurations of that LLM, temperature, and circuit. The two need not belong to the same configuration. \textbf{(b)} Retry overhead, measured as the number of generated response tokens including retries divided by the number in the accepted schedule. In both panels, boxes span the interquartile range and whiskers the remaining spread excluding outliers, with the individual runs overlaid as dots in the color of their distribution. Groups with fewer than three completed runs are drawn as tick marks instead of boxes.}
	\label{fig:run-distributions}
\end{figure*}

\subsubsection{Headline results}
\label{subsubsec:results-headline}
Our LLM-based compilers produce valid shuttling schedules on both trained architectures for circuits of up to 16 qubits. The rate of completed runs varies with the qubit count of the circuit and with the LLM. Of all runs, 36.6\,\% complete on the 7-qubit circuit \emph{4mod5-bdd\_287}, 4.9\,\% on the 10-qubit \emph{mini\_alu\_305}, and 0.3\,\% on the 16-qubit \emph{cnt3-5\_179}. Across circuits, Llama-3B completes 22.3\,\% of its runs, more than twice the rate of the next best LLM, Qwen-14B, at 9.3\,\%. These rates do not follow parameter count. Since ten runs are available per configuration, a second measure counts a configuration as \emph{successful} once one of its runs completes. By that measure 75\,\% of the configurations succeed on the 7-qubit circuit, 15\,\% on the 10-qubit one, and 1\,\% on the 16-qubit one. On the linear architecture of the 7-qubit circuit every LLM succeeds at both temperatures.

A schedule completed by an LLM is usually longer than the baseline schedule. The best LLM-generated schedule on the linear architecture, from Qwen-4B at $T=1.0$, requires 15\,\% fewer operations than the baseline, but averaged over the four completed runs of that configuration it requires 22\,\% more. Of the 113 successful configurations on the trained architectures, fourteen have a best run below their baseline, and five of those, all on the branched architecture, have a mean over their completed runs below it. Over individual runs, \autoref{fig:run-distributions} shows that 23 of the 506 completed runs, 4.5\,\%, fall below the baseline, and that a run requires 24\,\% more operations on average. Post-processing accounts for part of this: before it, every run sits further above its baseline, and 1.8\,\% of runs fall below it.

Compiling this way costs more than the classical compilers do. Counting the discarded responses as well, a schedule costs on average 2.6 times as many generated tokens, and at worst 8.9 times as many, as the accepted responses alone. Pooled over all completed runs, a schedule takes 96 minutes on average and 76 minutes at the median, ranging from 7 minutes to 6.3 hours.

The evidence for generalization on the three unseen layouts is preliminary. One of the ten runs of Llama at $T=1.0$ produces a complete schedule on the four-way layout of \autoref{fig:four-way-junction-trap}. Neither of the other two yields a schedule across the LLMs, temperatures, and circuits we tested, and on the theta layout no LLM executes a single gate.

\subsubsection{Linear and branched one-dimensional architectures}
\label{subsubsec:results-branched}
The results below are reported circuit by circuit, with both architectures for each. Three properties of the schedules are shared by all three circuits and are therefore reported once rather than repeated below: the proportion of each of the four shuttling operations in \autoref{subsubsec:results-composition}, the effect of the sampling temperature in \autoref{subsubsec:results-temperature}, and the reduction achieved by post-processing in \autoref{subsubsec:postprocessing}.

For \emph{4mod5-bdd\_287} on the linear architecture, whose results \autoref{tab:linear} lists, every LLM is successful at both temperatures, and 61 of the 100 runs complete in all. Four of the ten configurations produce schedules requiring fewer operations than the baseline, namely DeepSeek at both temperatures, Llama at $T=0.7$, and Qwen-4B at $T=1.0$. The best run is obtained by Qwen-4B at $T=1.0$, reducing the operation count by 15\,\%, though the mean over its four completed runs lies 22\,\% above the baseline. That LLM is also the most sensitive to $T$: at $T=0.7$ its best schedule contains 76\,\% more operations than the baseline, a difference of 91 percentage points, the largest of any LLM here. Post-processing accounts for little of either result: across these ten configurations it removes at most 5\,\% of the operations of a schedule, and at most 2\,\% of those of a best schedule. Across all LLMs, the generation took 2 to 26 retries, and valid schedules contain between 60{,}440 and 143{,}057 response tokens. With retries included, token counts grow by up to a factor of 6.2. The best Qwen-4B run also yields the smallest final token count, and its 26 retries make it the run with the highest token growth. Thus, the shortest schedule was reached through the largest number of attempts.

On the branched architecture, whose results \autoref{tab:branched-a} and \autoref{tab:branched-b} list, Llama produces valid schedules for all ten tested designs and outperforms the baseline in four of them, three at $T=1.0$ and one at $T=0.7$. Its best run relative to the baseline is obtained for $(5,1)$ at $T=0.7$, requiring 18.4\,\% fewer operations with ten retries. The shortest Llama schedule is a different one, 464 operations for $(2,3)$ at $T=1.0$, 5.1\,\% below its baseline after a single retry. Qwen-14B produces valid schedules for nine of the ten architecture designs, all except $(1,4)$, and surpasses the baseline for $(5,1)$ at both temperatures, by 7.2\,\% at $T=0.7$ and 20.7\,\% at $T=1.0$. The latter is the largest improvement over a baseline that we observe anywhere, although that configuration completed only one of its ten runs. Qwen-4B reaches valid schedules for eight architecture designs and outperforms the baseline for the same design as Qwen-14B, by 10.7\,\% at $T=1.0$.

Seven configurations across all LLMs beat their baseline on this circuit. Reliability and peak quality do not go together here: Llama completes every one of the ten designs but reaches at best $18.4$\,\% below its baseline, whereas Qwen-14B completes nine and reaches $20.7$\,\%. For five of the seven the mean over their completed runs also lies below the baseline. The remaining two are Llama's $(2,3)$ and $(3,1)$ at $T=1.0$, whose means lie 21.8\,\% and 14.1\,\% above it. None of the seven beats its baseline without the post-processing step, the raw counts lying between 0.6\,\% and 31.1\,\% above it. Gemma succeeds on seven architecture designs at $T=1.0$ and five at $T=0.7$, DeepSeek on five and one, and neither beats its baseline anywhere, coming no closer than 5.0\,\% and 4.0\,\% above it. Across the four LLMs other than Llama the best runs take between 23 and 31 retries, raising their token counts by up to a factor of 6.9.

For \emph{mini\_alu\_305} on the linear architecture, a single run out of 100 produces a complete schedule, by Qwen-4B at $T=1.0$, requiring 3{,}124 operations, 31\,\% more than the baseline. The accepted response holds 555{,}430 tokens, and the 33 retries needed along the way raise the total generated to 2.3 times as many. No run by any other LLM completes the circuit. The most progressed run reaches 161 gates of 263 for Llama, 123 for DeepSeek, 95 for Qwen-14B, and 75 for Gemma, against medians of 66, 35, 58, and 43.

On the branched architecture, out of 18 tested designs, Llama yields valid schedules for 15 at $T=0.7$ and 11 at $T=1.0$, three of which outperform the baseline. These are $(2,3)$ at $T=0.7$ and $(3,3)$ and $(2,3)$ at $T=1.0$, 3.9\,\%, 4.2\,\%, and 1.4\,\% below their baselines. They took 26, 53, and 31 retries, raising the generated tokens to 1.8, 3.8, and 2.3 times those required for the accepted schedule. All three improvements rest on the best-of-ten selection and post-processing. The means over the completed runs of those configurations lie 7.5\,\%, 2.3\,\%, and 8.5\,\% above their baselines, and the raw counts of the three best runs 21.7\,\%, 24.5\,\%, and 20.1\,\% above them. Qwen-4B also reaches a complete schedule, for $(5,1)$ at $T=1.0$, with 48.0\,\% more operations than the baseline. This run took 58 retries, raising the generated tokens to 3.5 times those required for the accepted schedule. Gemma, Qwen-14B, and DeepSeek are successful on no architecture design, reaching 126, 85, and 69 gates at best and 34, 28, and 13 at the median.

For \emph{cnt3-5\_179} on the linear architecture, none of the 100 runs completes the circuit. The most progressed run reaches 146 of the 230 gates for Llama, 118 for Qwen-4B, 98 for Qwen-14B, 88 for Gemma, and 63 for DeepSeek, against medians of 66, 66, 21, 16, and 16. Gemma is by far the most sensitive to $T$, reaching those 88 gates at $T=0.7$ and 43 at $T=1.0$. No other LLM shifts by more than 18 gates between the two temperatures.

On the branched architecture, out of 29 tested designs, Llama yields valid schedules for one at $T=0.7$ and two at $T=1.0$, and no other LLM for any. All three stay above the baseline, by 25.8\,\% for $(1,1)$ at $T=0.7$, by 31.4\,\% for $(1,1)$ at $T=1.0$, and by 15.9\,\% for $(3,1)$ at $T=1.0$. They took 27, 65, and 104 retries, raising the generated tokens to 2.2, 3.5, and 4.6 times those required for the accepted schedule. Llama's incomplete runs reach 201 gates of 230 at best. The most progressed run of the other LLMs reaches 140 for Qwen-4B, 64 for Qwen-14B, 56 for Gemma, and 49 for DeepSeek, against medians of 15, 12, 5, and 7.

For \emph{4mod5-bdd\_287}, one architecture design accounts for most of the improvements over the baseline. Among its ten branched designs, $(5,1)$ is the one the classical compiler handles worst, requiring 694 operations against 482 to 520 for the others, and it is where the fine-tuned LLMs are strongest. That correspondence does not hold in general: for \emph{mini\_alu\_305} the design handled worst by the baseline is $(6,1)$ and for \emph{cnt3-5\_179} it is $(1,1)$, and on neither does a schedule of the LLMs come out shorter. The stack height separates the designs more clearly than the junction distance does. The median schedule lies 17.6\,\% above the baseline at $s=1$ and 7.2\,\% below it at $s=5$, whereas over the four junction distances the median stays between 13.2 and 18.2\,\% above.

Counted over all five LLMs, all ten branched designs of \emph{4mod5-bdd\_287} yield a schedule, against 15 of 18 for \emph{mini\_alu\_305} and 2 of 29 for \emph{cnt3-5\_179}. Over the three circuits Llama accounts for 51 of the 113 configurations that produced a schedule and is the only LLM to complete one for all three circuits, while Qwen-4B accounts for 20 across two circuits and Qwen-14B, Gemma, and DeepSeek 20, 14, and 8 on \emph{4mod5-bdd\_287} alone. Eight of the fourteen configurations that beat their baseline are Llama's, two each are Qwen-4B's, Qwen-14B's, and DeepSeek's, and none is Gemma's.

\subsubsection{Schedule composition}
\label{subsubsec:results-composition}
The physical cost of a schedule is not determined by its operation count alone, since the primitives differ in duration and in the motional excitation they induce \cite{HilderPOSOLRMSP22}. The proportions in \autoref{tab:composition} are taken over the shuttling operations together with the gate executions. They are given per architecture, since pooling the two would average over distributions that differ.

On both architectures the generated schedules consist mostly of translations, the least demanding of the four primitives. Translations and gate executions together account for 78.6\,\% of the operations on the linear architecture and 90.4\,\% on the branched one. The remaining operations are separation, merge, and swap, which reshape a chain rather than move it, and their combined proportion is accordingly lower on the branched architecture. The difference is not an artifact of which configurations happened to complete. For each of the five LLMs at each temperature we compared the best schedule on the linear architecture with the best on the branched one, taking \emph{4mod5-bdd\_287}, the only circuit every LLM completes on both. In all ten pairs the combined proportion of separation, merge, and swap is lower on the branched architecture, falling from 19.4 to 26.4\,\% on the linear schedules to 11.2 to 16.3\,\% on the branched ones. The presence of stacks offers an explanation, since a qubit can be parked in one and later retrieved by translations alone. On the linear architecture the same reordering has to be routed through the gate segment, where the chain must be separated and merged again, and reversed by a swap if the order within a segment has to change. The physical cost of a schedule is therefore lower than its operation count suggests, and lower still on the branched architectures, where the same count is made up of less demanding operations.

\begin{table}[!t]
	\centering
	\setlength{\tabcolsep}{2pt}
	\scriptsize
	\caption{Proportion of each operation type in the complete schedules on the two trained architectures. The linear proportions are computed over 62 schedules and 51{,}401 operations, the branched ones over 444 schedules and 432{,}282 operations.}
	\label{tab:composition}
	\begin{tabular*}{\columnwidth}{@{\extracolsep{\fill}}lcc@{}}
		\toprule
		Operation & Linear & Branched \\
		\midrule
		\emph{Translate}      & 65.5\,\% & 75.9\,\% \\
		\emph{Execute Gate}   & 13.1\,\% & 14.5\,\% \\
		\emph{Separate}       &  8.6\,\% &  4.8\,\% \\
		\emph{Merge}          &  8.2\,\% &  4.4\,\% \\
		\emph{Swap}           &  4.7\,\% &  0.3\,\% \\
		\bottomrule
	\end{tabular*}
\end{table}

The swaps emitted on the branched architectures are not something the LLMs were shown during fine-tuning. Because the branched baseline \cite{KreppelMWHPSB24} eliminates swaps entirely, our training schedules for that architecture contain none, while those for the linear architecture do. The fine-tuned LLMs reproduce this on the linear architecture, where all 62 completed schedules contain swaps. On the branched architectures the LLMs nevertheless emit swaps, in 361 of the 444 completed schedules, a mean of 3.3 per schedule. They thus carry a behavior learned on the linear architecture over to one where it does not belong. Numerically the swaps are a negligible part of the 179 operations by which a schedule on the branched architecture exceeds its baseline in the mean. A preference-learning stage could decide between two treatments: suppressing the operation on architectures whose training data never uses it, or keeping it available for layouts where it helps.

\subsubsection{Unseen layouts}
\label{subsubsec:results-unseen}
The complete schedule on the four-way layout of \autoref{fig:four-way-junction-trap} comes from one of the ten runs of Llama at $T=1.0$, on \emph{4mod5-bdd\_287}, and needs 532 operations. The accepted response holds 68{,}000 tokens, and the 36 retries needed along the way raise the total generated to 3.4 times as many. Before post-processing the schedule holds 580 operations, and the run took 29 minutes. The nine runs alongside it reach a median of four gates. No baseline is available for comparison, since the classical compilers we use handle only the two trained architectures. At $T=0.7$, Llama reaches at most 12 of the 106 gates before the generation process becomes invalid. No other LLM completes a schedule on this layout. Over their twenty runs at the two temperatures, the most progressed reaches 52 gates for Qwen-14B, 33 for Qwen-4B, 16 for Gemma, and 4 for DeepSeek. The median over the same twenty runs is 7, 10, 0, and 4 gates.

Consistent with the trends seen on the trained architectures, compilation for the four-way layout gets harder as the number of qubits grows. On \emph{mini\_alu\_305} the most progressed run reaches 48 gates for Llama, 19 for Qwen-4B, 16 for Gemma, 15 for Qwen-14B, and none for DeepSeek, against medians of 8, 0, 0, 3, and 0. On \emph{cnt3-5\_179} it reaches 16 gates for Llama, 9 for Qwen-4B, 7 for Gemma, 5 for DeepSeek, and 4 for Qwen-14B, against medians of 7, 3, 0, 1, and 3.

The remaining two unseen layouts, the theta layout of \autoref{fig:theta-trap} and the multi-linear layout of \autoref{fig:multi-linear-trap}, yield no complete schedule. Every unseen layout was evaluated with 300 runs, ten for each of the five LLMs at two temperatures on three circuits. On the multi-linear layout the most progressed executes four of the 106 gates of the smallest circuit, and eight runs execute at least one gate. On the theta layout no LLM executes a gate.

The three layouts differ from the trained architectures in more than one way, and a single completed schedule cannot settle which difference matters. All three contain cycles, which the linear and branched architectures do not, so the presence of a cycle cannot explain why one of them nevertheless yielded a schedule. The two on which no complete schedule is produced have only three-way junctions, exactly as the branched training layouts do. In contrast, the one that succeeds has junctions of degree four, a degree absent from the training data. The single schedule produced there is preliminary evidence of cross-architecture generalization: an unseen layout is reachable, but the approach is not thereby layout-independent.

\begin{table*}[!t]
	\centering
	\setlength{\tabcolsep}{2pt}
	\scriptsize
	\caption{Runtime statistics for the five LLMs on the 7-qubit circuit \emph{4mod5-bdd\_287}, the only circuit every LLM compiles successfully, given separately for the two trained architectures. \emph{Runs} counts the completed runs, and \emph{Time} groups the minimum, mean, median, maximum, and standard deviation of their wall-clock runtime on a single A100. \emph{Rate} is the mean number of response tokens generated per second, counted over all responses including the discarded ones. It is given once because it differs by less than 5\,\% between the two architectures for every LLM except DeepSeek-7B, which reaches 69\,tok/s on the linear architecture and 57\,tok/s on the branched ones.}
	\label{tab:cost-models}
	\begin{tabular*}{\textwidth}{@{\extracolsep{\fill}}lrrrrrrrrrrrrr@{}}
		\toprule
		& \multicolumn{6}{c}{Linear architecture} & \multicolumn{6}{c}{Branched architecture} & \\
		\cmidrule(lr){2-7}\cmidrule(lr){8-13}
		& & \multicolumn{5}{c}{Time [min]} & & \multicolumn{5}{c}{Time [min]} & \\
		\cmidrule(lr){3-7}\cmidrule(lr){9-13}
		LLM & Runs & Min & Mean & Median & Max & SD & Runs & Min & Mean & Median & Max & SD & Rate [tok/s] \\
		\midrule
		Llama-3B     &  16 & 13 &  43 &  42 &  92 & 21 & 150 &  7 &  20 &  17 &  55 & 10 & 131 \\
		Qwen-4B      &   5 & 23 &  49 &  48 &  81 & 21 &  65 & 18 &  41 &  36 & 117 & 21 & 104 \\
		DeepSeek-7B  &  12 & 22 &  47 &  41 & 101 & 21 &   8 & 74 & 137 & 147 & 177 & 33 &  65 \\
		Qwen-14B     &  20 & 74 & 139 & 107 & 303 & 67 &  91 & 54 & 160 & 137 & 358 & 72 &  41 \\
		Gemma-27B    &   8 & 98 & 145 & 142 & 207 & 37 &  28 & 84 & 191 & 195 & 304 & 60 &  26 \\
		\bottomrule
	\end{tabular*}
\end{table*}

\subsubsection{Sampling temperature}
\label{subsubsec:results-temperature}
Across all 6{,}900 runs, the higher temperature yields slightly more completed runs: 7.9\,\% at $T=1.0$ against 6.8\,\% at $T=0.7$. The direction is consistent for four of the five LLMs, most clearly for Qwen-4B (6.7\,\% against 3.8\,\%), and reversed only for Qwen-14B (7.8\,\% against 8.3\,\%). The operation counts of the completed schedules are almost unaffected, with means differing by less than 3\,\% between the two temperatures on every circuit. Temperature therefore acts mainly on how often a valid schedule is found at all, rather than on how good it is. This matches \autoref{fig:run-distributions}, where the two temperatures of an LLM produce overlapping spreads of operation ratios, while the LLM and the circuit separate the groups.

\subsubsection{Post-processing}
\label{subsubsec:postprocessing}
The amount that the post-processing step of \autoref{subsec:inference} removes from a schedule depends strongly on the architecture. Averaged over the completed runs, it takes out 1.8\,\% of the raw operation count on the linear architecture and 13.8\,\% on the branched ones, with maxima of 8.3 and 34.4\,\% on individual schedules. The same separation holds per LLM: on the 7-qubit circuit the mean proportion removed ranges over the five LLMs between 0.6 and 3.3\,\% on the linear architecture and between 9.2 and 15.5\,\% on the branched ones. Because of this gap, post-processing decides the comparison against the baseline on the branched architectures, while on the linear architecture it barely influences the outcome. Of the fourteen configurations whose best run beats its baseline, ten are on the branched architectures and four on the linear one. On the raw counts only the latter four remain. The best of those four runs, from Qwen-4B, is 15\,\% below its baseline. Post-processing contributes little to that result, since the raw count of the same run is already 12.9\,\% below. No comparable step applies to the baselines, whose schedules are constructed rather than sampled and contain no redundancies to remove.

\subsubsection{Compilation cost}
\label{subsubsec:results-cost}
The practicality of the approach also depends on how long a compilation takes, and token counts do not measure that. Wall-clock runtimes do, but they are specific to the hardware used, which is why we report the two separately.

Apart from the LLM, two properties of a circuit drive its runtime. More gates take more time, and each gate takes longer on more qubits. Dividing a runtime by the gate count leaves only the second of these, which makes the mean time per gate comparable across circuits. Llama on the branched architecture is the only pairing that completes all three circuits. Its mean time per gate rises from 11 seconds on the 7-qubit circuit to 33 on the 10-qubit one and 72 on the 16-qubit one. Cost therefore outruns circuit size: the 16-qubit circuit has twice the gates of the 7-qubit one but takes almost fourteen times the runtime.

Holding the circuit fixed leaves the LLM and the architecture to vary. The runtime of the completed runs on the 7-qubit circuit is given in \autoref{tab:cost-models} for each LLM and architecture, with the LLMs ordered by parameter count. Their mean runtimes increase with that count on both architectures. The only exception is DeepSeek-7B, marginally faster than Qwen-4B on the linear architecture. Divided by the 106 gates a completed run executes, those means run from 11 to 108 seconds per gate on the branched architecture and from 24 to 82 seconds on the linear one. Generation speed falls as the parameter count rises, from 131 tokens per second for Llama-3B down to 26 for Gemma-27B. For every LLM and architecture the slowest run lies well above the mean. Qwen-14B on the branched architecture shows the widest gap, at 358 minutes against a mean of 160 minutes.

Our evaluation consumed approximately 12{,}900 GPU-hours, of which 94\,\% went to runs that stopped without a complete schedule. Pooled over all circuits and LLMs, such a run costs more than a completed one, 113 minutes on average against 96, because it works through the retry loop to its limit first. The baseline compilers \cite{Wagner22, KreppelMWHPSB24} compile the same instances on a single CPU core in milliseconds, four to six orders of magnitude faster. Shorter schedules remain worth having even at that price, since a schedule that is 15\,\% shorter raises the probability that the circuit runs to completion on the device before decoherence sets in, which can outweigh minutes spent compiling it.

\subsubsection{Schedules shorter than the baselines}
\label{subsubsec:beating}
On fourteen configurations the fine-tuned LLMs produce shorter schedules than the classical compilers whose output they were trained on. Two properties of the training data and one of the baselines make this possible. First, each example covers a single gate and maps a trap state and the pending gates to the operations that follow. The LLM therefore learns a local policy over trap states rather than a particular schedule. Second, the examples are generated for random circuits, whereas the three test circuits come from the circuit library \cite{Zhou19}, so no test schedule appears in the training data. Third, since the baselines are heuristics rather than optimal solvers, a policy fitted to their behavior is not bound by their output on any individual circuit.

The LLMs have learned an approximation of the classical routing strategy that varies from run to run. That spread is visible in \autoref{fig:run-distributions}: most completed runs need more operations than the baseline, and the few that reach below it are the ones best-of-ten sampling selects.

\subsection{Limitations}
\label{subsec:limitations}
The two baselines \cite{Wagner22, KreppelMWHPSB24} are highly optimized heuristics, so no comparison in this paper is against an optimum. Since the number of legal operation sequences grows exponentially with the length of a schedule, no exact solver is available for the segmented trap with chain-level operations, and the optimum is consequently unknown. We evaluated 690 configurations, five LLMs at two temperatures on each of 69 pairings of a circuit with a trained architecture or an unseen layout, and 600 of them have a baseline. On fourteen of these our best schedules improve on the heuristic, but how much further an optimal schedule would go remains open.

A complete schedule was produced by 114 of the 690 configurations. Measured as the proportion of a circuit's gates, the furthest run among the remainder comes from Llama at $T=1.0$ on the branched design $(3,2)$ of \emph{mini\_alu\_305}, where it executes 247 of the 263 gates. A schedule is usable only once every gate is scheduled. The progress shown in \autoref{fig:results-heatmap} therefore serves to locate the point in a circuit at which generation breaks down.

Compiling circuits of more than 16 qubits would require training data at those sizes, and such data alone is unlikely to suffice: circuits of 16 qubits are already covered, and only 0.3\,\% of the runs on them complete. The five LLMs span 3 to 27 billion parameters, and within that range the completion rate does not follow model size. Larger LLMs were beyond our GPU budget and remain untested.

Two limitations follow from the design of the compiler. First, inference proceeds one first-layer gate at a time, so the LLM has no information about how the trap state it leaves behind will affect the gates that follow. A sequence that is short for the current gate can therefore leave a state from which the next gates need many further operations. This becomes more likely as circuits grow deeper, and is a plausible contributor to the runs that stop before the last gate. Second, the LLMs are trained on the schedules a classical compiler produced, without any comparison to alternative sequences for the same trap state, so training rewards validity rather than length.

Two further limitations concern the scope of our results. First, of the three unseen layouts we evaluated, only one yielded a schedule, with one LLM in a single run. A new trap family still needs a classical compiler of its own to generate the training data, though one such compiler then serves every configuration in that family. Removing that dependence would require training data from a source other than an existing compiler, or a transfer between trap families. Second, the comparison against the baselines is favorable only in the best-of-ten reading, and on the branched architectures it depends strongly on the post-processing step. The best-of-ten qualification is needed because the runs vary, and becomes unnecessary once every run reaches the operation count that the best of ten reaches currently.

\section{Conclusion and Outlook}
\label{sec:conclusion}

We have presented the first shuttling compilers for trapped-ion quantum computers based on LLMs. We fine-tuned five open-weight LLMs from four families on schedules from classical compilers for linear \cite{Wagner22} and branched one-dimensional \cite{KreppelMWHPSB24} trap architectures, and evaluated them on benchmark circuits of up to 16 qubits.

On both training architectures the fine-tuned LLMs generate valid schedules, the more of them the fewer qubits a circuit has. Of the 600 combinations of LLM, sampling temperature, circuit, and trained architecture, 113 produced a valid schedule. On fourteen of these configurations, 12\,\%, the best of the ten runs needs up to 21\,\% fewer operations than the baseline heuristics \cite{Wagner22, KreppelMWHPSB24} after a rule-based post-processing step. Those heuristics are the compilers that generated the training data. For five of the fourteen configurations, all on the branched architecture, the mean over the completed runs is below the baseline as well, while four others, all on the linear architecture, already beat it before post-processing. Of three previously unseen layouts, one yields a valid schedule in a single run of one LLM and the other two yield none, preliminary evidence that a learned compiler can transfer to a layout it was not trained on. The proportion of runs that complete does not follow the size of the LLM: the smallest, with 3 billion parameters, completes 22\,\% of its runs, more than twice the rate of any larger one. Pooled over all completed runs, a schedule takes 96 minutes on average against milliseconds for the baselines.

Our work shows that shuttling schedules can be learned from examples alone, without hand-coded routing logic. Matching a classical compiler in schedule quality, compile time, and the circuit sizes it can handle is the next step.

We see three directions for future work. (i) We plan to add a preference-learning stage based on Direct Preference Optimization (DPO) \cite{RafailovSMEMF23}. Candidate sequences would be ranked by a bounded-depth lookahead over the following gates rather than by their own length. Such a ranking would also settle whether the swaps our LLMs emit on the branched architectures help, since a sequence containing one would be scored against a sequence without it. (ii) We intend to follow this with Group Relative Policy Optimization (GRPO) \cite{ShaoWZXS24}, which scores a group of candidates for the same trap state against one another under a rule-based reward that penalizes invalid schedules and rewards short valid ones. Both stages require generating many candidates per gate and will benefit from more compute than the eight H100 GPUs available here. (iii) Finally, we consider a control experiment that trains from scratch equally important. We would take one of the LLMs we fine-tuned, instantiate it from its own configuration and tokenizer but with randomly initialized weights, and train and evaluate it on the same data and circuits. That would separate the contribution of pre-training from that of the transformer architecture and the fine-tuning data. Our results show that fine-tuning is necessary, but not whether pre-training is. Our learning rate, schedule, and number of epochs were chosen for an LLM that starts from pre-trained weights. Both the randomly initialized LLM and the fine-tuned one would need values chosen for them. Under ours, a failure could mean either that pre-training is needed or that the training procedure never suited a randomly initialized LLM.

\subsection*{Acknowledgments}
\label{sec:acknowledgments}

We acknowledge funding by the German Federal Ministry of Research, Technology and Space (BMFTR) within the project IQuAn, ATIQ and SYNQ, and the DFG SPP2514. We thank the data center (ZDV) of the Johannes Gutenberg University for the use of their Mogon NHR and Mogon KI clusters for evaluating our LLMs.

\subsection*{Conflicts of Interest}

The authors declare no conflicts of interest.

\subsection*{Data Availability Statement}

The datasets, training scripts, and trained models that support the findings of this study are available from the authors upon request.

\bibliographystyle{quantum}
\bibliography{references}

@inproceedings{AshSakiTG22,
  author       = {Abdullah Ash{-}Saki and Rasit Onur Topaloglu and Swaroop Ghosh},
  title        = {Muzzle the {S}huttle: {E}fficient {C}ompilation for {M}ulti{-T}rap {T}rapped{-I}on {Q}uantum {C}omputers},
  booktitle    = {Des. Automat. Test Europe Conf. Exhib. (DATE)},
  pages        = {322--327},
  publisher    = {IEEE},
  address      = {Antwerp, Belgium},
  year         = {Mar. 14--23, 2022},
  doi          = {10.23919/DATE54114.2022.9774619},
  url          = {https://www.doi.org/10.23919/DATE54114.2022.9774619},
  eprint       = {2111.07961},
}

@misc{Axolotl23,
  title        = {Axolotl: {O}pen {S}ource {LLM} {P}ost{-T}raining},
  author       = {{Axolotl maintainers and contributors}},
  year         = {2023},
  month        = may,
  url          = {https://github.com/axolotl-ai-cloud/axolotl},
  license      = {Apache-2.0},
}

@article{BachSY26,
  author       = {Bao G. Bach and Ilya Safro and Ed Younis},
  title        = {Efficient {C}ompilation for {S}huttling {T}rapped{-I}on {M}achines via the {P}osition {G}raph {A}rchitectural {A}bstraction},
  journal      = {{ACM} Trans. Quantum Comput. (TQC)},
  volume       = {7},
  number       = {4},
  pages        = {29:1--29:33},
  year         = {2026},
  month        = aug,
  doi          = {10.1145/3831246},
  url          = {https://www.doi.org/10.1145/3831246},
  eprint       = {2501.12470},
}

@misc{BergholmISG+22,
  author       = {Ville Bergholm and others},
  title        = {Penny{L}ane: {A}utomatic differentiation of hybrid quantum{-}classical computations},
  publisher    = {arXiv},
  archivePrefix= {arXiv},
  primaryClass = {quant-ph},
  year         = {2022},
  month        = jul,
  doi          = {10.48550/arXiv.1811.04968},
  url          = {https://www.doi.org/10.48550/arXiv.1811.04968},
  eprint       = {1811.04968},
}

@article{BlakestadODWBLW11,
  author       = {R. B. Blakestad and C. Ospelkaus and A. P Van{D}evender and J. H Wesenberg and M. J Biercuk and D. Leibfried and D. J Wineland},
  title        = {Near{-}ground{-}state transport of trapped{-}ion qubits through a multidimensional array},
  journal      = {Phys. Rev. A},
  volume       = {84},
  number       = {3},
  pages        = {032314},
  year         = {2011},
  month        = sep,
  doi          = {10.1103/PhysRevA.84.032314},
  url          = {https://www.doi.org/10.1103/PhysRevA.84.032314},
}

@article{BlakestadOVABLW09,
  author       = {R. B. Blakestad and C. Ospelkaus and A.P. Van{D}evender and J.M. Amini and J. Britton and D. Leibfried and D.J. Wineland},
  title        = {High{-F}idelity {T}ransport of {T}rapped{-I}on {Q}ubits through an $\mathbf{X}${-J}unction {T}rap {A}rray},
  journal      = {Phys. Rev. Lett.},
  volume       = {102},
  number       = {15},
  pages        = {153002},
  year         = {2009},
  month        = apr,
  doi          = {10.1103/PhysRevLett.102.153002},
  url          = {https://www.doi.org/10.1103/PhysRevLett.102.153002},
  eprint       = {0901.0533},
}

@article{BowlerGLTHJHLW12,
  author       = {R. Bowler and J. Gaebler and Y. Lin and T. R. Tan and D. Hanneke and J. D. Jost and J. P. Home and D. Leibfried and D. J. Wineland},
  title        = {Coherent {D}iabatic {I}on {T}ransport and {S}eparation in a {M}ultizone {T}rap {A}rray},
  journal      = {Phys. Rev. Lett.},
  volume       = {109},
  number       = {8},
  pages        = {080502},
  year         = {2012},
  month        = aug,
  doi          = {10.1103/PhysRevLett.109.080502},
  url          = {https://www.doi.org/10.1103/PhysRevLett.109.080502},
  eprint       = {1206.0780},
}

@inproceedings{Brown+20,
  author       = {Tom B. Brown and others},
  title        = {Language models are few-shot learners},
  booktitle    = {34th Int. Conf. Neural Inf. Process. Syst. (NeurIPS)},
  pages        = {1877--1901},
  publisher    = {Curran Associates Inc.},
  address      = {Virtual event},
  year         = {Dec. 6--12, 2020},
  url          = {https://dl.acm.org/doi/abs/10.5555/3495724.3495883},
  eprint       = {2005.14165},
}

@article{ChangJCHL25,
  author       = {Che{-}Ming Chang and Jie{-}Hong Roland Jiang and Dah{-}Wei Chiou and Ting Hsu and Guin-Dar Lin},
  title        = {Quantum {C}ircuit {C}ompilation for {T}rapped{-I}on {P}rocessors {W}ith the {D}rive{-T}hrough {A}rchitecture},
  journal      = {{IEEE} Trans. Quantum Eng. (TQE)},
  volume       = {6},
  pages        = {1--14},
  year         = {2025},
  month        = mar,
  doi          = {10.1109/TQE.2025.3548423},
  url          = {https://www.doi.org/10.1109/TQE.2025.3548423},
}

@misc{ChenXZG16,
  author       = {Tianqi Chen and Bing Xu and Chiyuan Zhang and Carlos Guestrin},
  title        = {Training {D}eep {N}ets with {S}ublinear {M}emory {C}ost},
  publisher    = {arXiv},
  archivePrefix= {arXiv},
  primaryClass = {cs.LG},
  year         = {2016},
  month        = apr,
  doi          = {10.48550/arXiv.1604.06174},
  url          = {https://www.doi.org/10.48550/arXiv.1604.06174},
  eprint       = {1604.06174},
}

@book{CirqDevelopers25, 
  author       = {{Cirq Developers}},
  title        = {Cirq},
  abstractNote = {Python package for writing, manipulating, and running quantum circuits on quantum computers and simulators},
  publisher    = {Zenodo},
  year         = {2025},
  month        = aug,
  doi          = {10.5281/ZENODO.4062499},
  url          = {https://www.doi.org/10.5281/ZENODO.4062499},
}

@inproceedings{CowtanDDKSS19,
  author       = {Alexander Cowtan and Silas Dilkes and Ross Duncan and Alexandre Krajenbrink and Will Simmons and Seyon Sivarajah},
  title        = {On the {Q}ubit {R}outing {P}roblem},
  booktitle    = {14th Conf. Theory Quantum Comput. Commun. Cryptography (TQC)},
  pages        = {5:1--5:32},
  publisher    = {Schloss Dagstuhl -- Leibniz-Zentrum für Informatik},
  address      = {Univ. of Maryland, College Park, MD, USA},
  year         = {June 3--5, 2019},
  doi          = {10.4230/LIPIcs.TQC.2019.5},
  url          = {https://www.doi.org/10.4230/LIPIcs.TQC.2019.5},
  eprint       = {1902.08091},
}

@misc{CrossBSG17,
  author       = {Andrew W. Cross and Lev S. Bishop and John A. Smolin and Jay M. Gambetta},
  title        = {Open {Q}uantum {A}ssembly {L}anguage},
  publisher    = {arXiv},
  archivePrefix= {arXiv},
  primaryClass = {quant-ph},
  year         = {2017},
  month        = jul,
  doi          = {10.48550/arXiv.1707.03429},
  url          = {https://www.doi.org/10.48550/arXiv.1707.03429},
  eprint       = {1707.03429},
}

@article{CrossJADBHRSSGJ22,
  author       = {Andrew Cross and Ali Javadi{-}Abhari and Thomas Alexander and Niel De Beaudrap and Lev S. Bishop and Steven Heidel and Colm A. Ryan and Prasahnt Sivarajah and John Smolin and Jay M. Gambetta and Blake R. Johnson},
  title        = {Open{QASM} 3: {A} {B}roader and {D}eeper {Q}uantum {A}ssembly {L}anguage},
  journal      = {{ACM} Trans. Quantum Comput. (TQC)},
  volume       = {3},
  number       = {3},
  pages        = {12:1--12:50},
  year         = {2022},
  month        = sep,
  doi          = {10.1145/3505636},
  url          = {https://www.doi.org/10.1145/3505636},
  eprint       = {2104.14722},
}

@article{DaiBR24,
  author       = {Weining Dai and Kevin A. Brown and Thomas G. Robertazzi},
  title        = {Advanced {S}huttle {S}trategies for {P}arallel {QCCD} {A}rchitectures},
  journal      = {{IEEE} Trans. Quantum Eng. (TQE)},
  volume       = {5},
  pages        = {1--18},
  year         = {2024},
  month        = jun,
  doi          = {10.1109/TQE.2024.3408757},
  url          = {https://www.doi.org/10.1109/TQE.2024.3408757},
}

@inproceedings{Dao24,
  author       = {Tri Dao},
  title        = {Flash{A}ttention{-}2: {F}aster {A}ttention with {B}etter {P}arallelism and {W}ork {P}artitioning},
  booktitle    = {12th Int. Conf. Learn. Representations (ICLR)},
  pages        = {1--14},
  publisher    = {OpenReview},
  address      = {Vienna, Austria},
  year         = {May 7--11, 2024},
  url          = {https://openreview.net/forum?id=mZn2Xyh9Ec},
  eprint       = {2307.08691},
}

@misc{Deepseek-AI24,
  author       = {{DeepSeek-AI}},
  title        = {Deep{S}eek {LLM}: {S}caling {O}pen{-S}ource {L}anguage {M}odels with {L}ongtermism},
  publisher    = {arXiv},
  archivePrefix= {arXiv},
  primaryClass = {cs.CL},
  year         = {2024},
  month        = jan,
  doi          = {10.48550/arXiv.2401.02954},
  url          = {https://www.doi.org/10.48550/arXiv.2401.02954},
  eprint       = {2401.02954},
}

@misc{Deepseek-AI24_2,
  author       = {{DeepSeek-AI}},
  title        = {Deep{S}eek{-V3} {T}echnical {R}eport},
  publisher    = {arXiv},
  archivePrefix= {arXiv},
  primaryClass = {cs.CL},
  year         = {2024},
  month        = dec,
  doi          = {10.48550/arXiv.2412.19437},
  url          = {https://www.doi.org/10.48550/arXiv.2412.19437},
  eprint       = {2412.19437},
}

@inproceedings{DettmersPHZ23,
  author       = {Tim Dettmers and Artidoro Pagnoni and Ari Holtzman and Luke Zettlemoyer},
  title        = {Q{L}o{RA}: {E}fficient {F}inetuning of {Q}uantized {LLM}s},
  booktitle    = {37th Int. Conf. Neural Inf. Process. Syst. (NeurIPS)},
  pages        = {10088--10115},
  publisher    = {Curran Associates Inc.},
  address      = {New Orleans, LA, USA},
  year         = {Dec. 10--16, 2023},
  url          = {https://dl.acm.org/doi/10.5555/3666122.3666563},
  eprint       = {2305.14314},
}

@misc{DurandauBSPMB26,
  author       = {Jonathan Durandau and Charles{-}Antoine Brunet and Ferdinand Schmidt{-}Kaler and Ulrich Poschinger and Frédéric Mailhot and Yves Bérubé{-}Lauzière},
  title        = {Heuristics for {S}huttling {S}equence {O}ptimization for a {L}inear {S}egmented {T}rapped{-I}on {Q}uantum {C}omputer},
  publisher    = {arXiv},
  archivePrefix= {arXiv},
  primaryClass = {quant-ph},
  year         = {2026},
  month        = mar,
  doi          = {10.48550/arXiv.2603.05464},
  url          = {https://www.doi.org/10.48550/arXiv.2603.05464},
  eprint       = {2603.05464},
}

@article{DurandauWMBSPB23,
  author       = {Jonathan Durandau and Janis Wagner and Frédéric Mailhot and Charles{-}Antoine Brunet and Ferdinand Schmidt{-}Kaler and Ulrich Poschinger and Yves Bérubé{-}Lauzière},
  title        = {Automated {G}eneration of {S}huttling {S}equences for a {L}inear {S}egmented {I}on {T}rap {Q}uantum {C}omputer},
  journal      = {Quantum},
  volume       = {7},
  pages        = {1175},
  year         = {2023},
  month        = nov,
  doi          = {10.22331/q-2023-11-08-1175},
  url          = {https://www.doi.org/10.22331/q-2023-11-08-1175},
  eprint       = {2208.04881},
}

@inproceedings{FanGL22,
  author       = {Hongxiang Fan and Ce Guo and Wayne Luk},
  title        = {Optimizing quantum circuit placement via machine learning},
  booktitle    = {59th {ACM/IEEE} Des. Automat. Conf. (DAC)},
  pages        = {19--24},
  publisher    = {ACM},
  address      = {San Francisco, CA, USA},
  year         = {July 10--14, 2022},
  doi          = {10.1145/3489517.3530403},
  url          = {https://www.doi.org/10.1145/3489517.3530403},
}

@misc{GemmaTeam25,
  author       = {{Gemma Team}},
  title        = {Gemma 3 {T}echnical {R}eport},
  publisher    = {arXiv},
  archivePrefix= {arXiv},
  primaryClass = {cs.CL},
  year         = {2025},
  month        = mar,
  doi          = {10.48550/arXiv.2503.19786},
  url          = {https://www.doi.org/10.48550/arXiv.2503.19786},
  eprint       = {2503.19786},
}

@article{GuoYZ+25,
  author       = {Daya Guo and Dejian Yang and Haowei Zhang and others},
  title        = {Deep{S}eek{-R1} incentivizes reasoning in {LLM}s through reinforcement learning},
  journal      = {Nature},
  volume       = {645},
  number       = {8081},
  pages        = {633--638},
  year         = {2025},
  month        = sep,
  doi          = {10.1038/s41586-025-09422-z},
  url          = {https://www.doi.org/10.1038/s41586-025-09422-z},
}

@article{HensingerOSHYADMR06,
  author       = {W. K. Hensinger and S. Olmschenk and D. Stick and D. Hucul and M. Yeo and M. Acton and L. Deslauriers and J. Rabchuk and C. Monroe},
  title        = {T{-}junction ion trap array for two-dimensional ion shuttling, storage, and manipulation},
  journal      = {Appl. Phys. Lett.},
  volume       = {88},
  number       = {3},
  pages        = {034101},
  year         = {2006},
  month        = jan,
  doi          = {10.1063/1.2164910},
  url          = {https://www.doi.org/10.1063/1.2164910},
  eprint       = {quant-ph/0508097},
}

@article{HilderPOSOLRMSP22,
  author       = {Janine Hilder and Daniel Pijn and Oleksiy Onishchenko and Alexander Stahl and Maximilian Orth and Björn Lekitsch and Andrea Rodriguez{-}Blanco and Markus Müller and Ferdinand Schmidt{-}Kaler and Ulrich Poschinger},
  title        = {Fault{-T}olerant {P}arity {R}eadout on a {S}huttling{-B}ased {T}rapped{-I}on {Q}uantum {C}omputer},
  journal      = {Phys. Rev. X},
  volume       = {12},
  number       = {1},
  pages        = {011032},
  year         = {2022},
  month        = feb,
  doi          = {10.1103/PhysRevX.12.011032},
  url          = {https://www.doi.org/10.1103/PhysRevX.12.011032},
  eprint       = {2107.06368},
}

@inproceedings{HuSWALWWC22,
  author       = {Edward J. Hu and Yelong Shen and Phillip Wallis and Zeyuan Allen-Zhu and Yuanzhi Li and Shean Wang and Lu Wang and Weizhu Chen},
  title        = {Lo{RA}: {L}ow{-R}ank {A}daptation of {L}arge {L}anguage {M}odels},
  booktitle    = {10th Int. Conf. Learn. Representations (ICLR)},
  pages        = {1--13},
  publisher    = {OpenReview},
  address      = {Virtual event},
  year         = {Apr. 25--29, 2022},
  url          = {https://openreview.net/forum?id=nZeVKeeFYf9},
  eprint       = {2106.09685},
}

@misc{JavadiAbhariTKWLGMNBCJG24,
  author       = {Ali Javadi{-}Abhari and Matthew Treinish and Kevin Krsulich and Christopher J. Wood and Jake Lishman and Julien Gacon and Simon Martiel and Paul D. Nation and Lev S. Bishop and Andrew W. Cross and Blake R. Johnson and Jay M. Gambetta},
  title        = {Quantum computing with {Q}iskit},
  publisher    = {arXiv},
  archivePrefix= {arXiv},
  primaryClass = {quant-ph},
  year         = {2024},
  month        = may,
  doi          = {10.48550/arXiv.2405.08810},
  url          = {https://www.doi.org/10.48550/arXiv.2405.08810},
  eprint       = {2405.08810},
}

@article{KaushalLSHPSBMSP20,
  author       = {V. Kaushal and B. Lekitsch and A. Stahl and J. Hilder and D. Pijn and C. Schmiegelow and A. Bermudez and M. Müller and F. Schmidt{-}Kaler and U. Poschinger},
  title        = {Shuttling{-}based trapped{-}ion quantum information processing},
  journal      = {{AVS} Quantum Sci.},
  volume       = {2},
  number       = {1},
  pages        = {014101},
  year         = {2020},
  month        = mar,
  doi          = {10.1116/1.5126186},
  url          = {https://www.doi.org/10.1116/1.5126186},
  eprint       = {1912.04712},
}

@article{KielpinskiMW02,
  author       = {D. Kielpinski and C. Monroe and D. J. Wineland},
  title        = {Architecture for a large{-}scale ion{-}trap quantum computer},
  journal      = {Nature},
  volume       = {417},
  number       = {6890},
  pages        = {709--711},
  year         = {2002},
  month        = jun,
  doi          = {10.1038/nature00784},
  url          = {https://www.doi.org/10.1038/nature00784},
}

@inproceedings{KorthikantiCLMASC23,
  author       = {Vijay Korthikanti and Jared Casper and Sangkug Lym and Lawrence McAfee and Michael Andersch and Mohammad Shoeybi and Bryan Catanzaro},
  title        = {Reducing {A}ctivation {R}ecomputation in {L}arge {T}ransformer {M}odels},
  booktitle    = {6th Conf. Mach. Learn. Syst. (MLSys)},
  pages        = {341--353},
  publisher    = {Curran Associates Inc.},
  address      = {Miami Beach, FL, USA},
  year         = {June 4--8, 2023},
  url          = {https://proceedings.mlsys.org/paper_files/paper/2023/file/80083951326cf5b35e5100260d64ed81-Paper-mlsys2023.pdf},
  eprint       = {2205.05198},
}

@article{KreppelMOWHPSB23,
  author       = {Fabian Kreppel and Christian Melzer and Diego Olvera Millán and Janis Wagner and Janine Hilder and Ulrich Poschinger and Ferdinand Schmidt{-}Kaler and André Brinkmann},
  title        = {Quantum {C}ircuit {C}ompiler for a {S}huttling{-B}ased {T}rapped{-I}on {Q}uantum {C}omputer},
  journal      = {Quantum},
  volume       = {7},
  pages        = {1176},
  year         = {2023},
  month        = nov,
  doi          = {10.22331/q-2023-11-08-1176},
  url          = {https://www.doi.org/10.22331/q-2023-11-08-1176},
  eprint       = {2207.01964},
}

@inproceedings{KreppelMWHPSB24,
  author       = {Fabian Kreppel and Christian Melzer and Janis Wagner and Janine Hilder and Ulrich Poschinger and Ferdinand Schmidt-Kaler and André Brinkmann},
  title        = {Shuttling {C}ompiler for a {T}rapped{-I}on {Q}uantum {C}omputer {A}rchitecture with {J}unctions},
  booktitle    = {{IEEE} Int. Conf. Quantum Comput. Eng. (QCE)},
  volume       = {1},
  pages        = {1065--1076},
  publisher    = {IEEE},
  address      = {Montreal, Canada},
  year         = {Sept. 15--20, 2024},
  doi          = {10.1109/QCE60285.2024.00126},
  url          = {https://www.doi.org/10.1109/QCE60285.2024.00126},
}

@misc{KunduLWGM24,
  author       = {Achintya Kundu and Rhui Dih Lee and Laura Wynter and Raghu Kiran Ganti and Mayank Mishra},
  title        = {Enhancing {T}raining {E}fficiency {U}sing {P}acking with {F}lash {A}ttention},
  publisher    = {arXiv},
  archivePrefix= {arXiv},
  primaryClass = {cs.LG},
  year         = {2024},
  month        = jul,
  doi          = {10.48550/arXiv.2407.09105},
  url          = {https://www.doi.org/10.48550/arXiv.2407.09105},
  eprint       = {2407.09105},
}

@inproceedings{KwonLZSZYGZS23,
  author       = {Woosuk Kwon and Zhuohan Li and Siyuan Zhuang and Ying Sheng and Lianmin Zheng and Cody Hao Yu and Joseph E. Gonzalez and Hao Zhang and Ion Stoica},
  title        = {Efficient {M}emory {M}anagement for {L}arge {L}anguage {M}odel {S}erving with {P}aged{A}ttention},
  booktitle    = {29th {ACM} Symp. Operating Syst. Princ. (SOSP)},
  pages        = {611--626},
  publisher    = {ACM},
  address      = {Koblenz, Germany},
  year         = {Oct. 23--26, 2023},
  doi          = {10.1145/3600006.3613165},
  url          = {https://www.doi.org/10.1145/3600006.3613165},
  eprint       = {2309.06180},
}

@article{LeeJPJKC21,
  author       = {Minjae Lee and Junho Jeong and Yunjae Park and Changhyun Jung and Taehyun Kim and Dong{-}il Cho},
  title        = {Ion shuttling method for long{-}range shuttling of trapped ions in {MEMS-}fabricated ion traps},
  journal      = {Jpn. J. Appl. Phys.},
  volume       = {60},
  number       = {2},
  pages        = {027004},
  year         = {2021},
  month        = feb,
  doi          = {10.35848/1347-4065/abdabb},
  url          = {https://www.doi.org/10.35848/1347-4065/abdabb},
}

@article{LekitschWFMDWH17,
  author       = {Bjoern Lekitsch and Sebastian Weidt and Austin G. Fowler and Klaus Mølmer and Simon J. Devitt and Christof Wunderlich and Winfried K. Hensinger},
  title        = {Blueprint for a microwave trapped ion quantum computer},
  journal      = {Sci. Adv.},
  volume       = {3},
  number       = {2},
  pages        = {e1601540},
  year         = {2017},
  month        = feb,
  doi          = {10.1126/sciadv.1601540},
  url          = {https://www.doi.org/10.1126/sciadv.1601540},
  eprint       = {1508.00420},
}

@inproceedings{LiDX19,
  author       = {Gushu Li and Yufei Ding and Yuan Xie},
  title        = {Tackling the {Q}ubit {M}apping {P}roblem for {NISQ-E}ra {Q}uantum {D}evices},
  booktitle    = {24th Int. Conf. Archit. Support Program. Lang. Operating Syst. (ASPLOS)},
  pages        = {1001--1014},
  publisher    = {ACM},
  address      = {Providence, RI, USA},
  year         = {Apr. 13--17, 2019},
  doi          = {10.1145/3297858.3304023},
  url          = {https://www.doi.org/10.1145/3297858.3304023},
  eprint       = {1809.02573},
}

@inproceedings{LiuZA24,
  author       = {Hao Liu and Matei Zaharia and Pieter Abbeel},
  title        = {Ring {A}ttention with {B}lockwise {T}ransformers for {N}ear{-I}nfinite {C}ontext},
  booktitle    = {12th Int. Conf. Learn. Representations (ICLR)},
  pages        = {1--17},
  publisher    = {OpenReview},
  address      = {Vienna, Austria},
  year         = {May 7--11, 2024},
  url          = {https://openreview.net/forum?id=fXugVDtCQO},
  eprint       = {2310.01889},
}

@inproceedings{LoshchilovH19,
  author       = {Ilya Loshchilov and Frank Hutter},
  title        = {Decoupled {W}eight {D}ecay {R}egularization},
  booktitle    = {7th Int. Conf. Learn. Representations (ICLR)},
  pages        = {1--18},
  publisher    = {OpenReview},
  address      = {New Orleans, LA, USA},
  year         = {May 6--9, 2019},
  url          = {https://openreview.net/forum?id=Bkg6RiCqY7},
  eprint       = {1711.05101},
}

@repository{MetaAI24,
  author       = {{Meta AI}},
  title        = {Model {I}nformation {LLaMa} 3.2},
  year         = {2024},
  month        = oct,
  note         = {{G}it{H}ub repository},
  code         = {https://github.com/meta-llama/llama-models/blob/main/models/llama3_2/MODEL_CARD.md},
  commit       = {4a9ff6a},
}

@repository{MetaAI25,
  author       = {{Meta AI}},
  title        = {Model {I}nformation {LLaMa} 4},
  year         = {2025},
  month        = apr,
  note         = {{G}it{H}ub repository},
  code         = {https://github.com/meta-llama/llama-models/blob/main/models/llama4/MODEL_CARD.md},
  commit       = {038acb9},
}

@article{MokhberiSW17,
  author       = {Arezoo Mokhberi and Roman Schmied and Stefan Willitsch},
  title        = {Optimised surface{-}electrode ion{-}trap junctions for experiments with cold molecular ions},
  journal      = {New J. Phys.},
  volume       = {19},
  number       = {4},
  pages        = {043023},
  year         = {2017},
  month        = apr,
  doi          = {10.1088/1367-2630/aa6918},
  url          = {https://www.doi.org/10.1088/1367-2630/aa6918},
  eprint       = {1701.06408},
}

@article{MolaviXCTA26,
  author       = {Abtin Molavi and Amanda Xu and Ethan Cecchetti and Swamit Tannu and Aws Albarghouthi},
  title        = {Generating {C}ompilers for {Q}ubit {M}apping and {R}outing},
  journal      = {Proc. {ACM} Program. Lang. (PACMPL)},
  volume       = {10},
  number       = {POPL},
  pages        = {2265--2294},
  year         = {2026},
  doi          = {10.1145/3776720},
  url          = {https://www.doi.org/10.1145/3776720},
  eprint       = {2508.10781},
}

@article{MosesB+23,
  author       = {S. A. Moses and C. H. Baldwin and others},
  title        = {A {R}ace{-T}rack {T}rapped-{I}on {Q}uantum {P}rocessor},
  journal      = {Phys. Rev. X},
  volume       = {13},
  number       = {4},
  pages        = {041052},
  year         = {2023},
  month        = dec,
  doi          = {10.1103/PhysRevX.13.041052},
  url          = {https://www.doi.org/10.1103/PhysRevX.13.041052},
  eprint       = {2305.03828},
}

@inproceedings{MuraliDBM20,
  author       = {Prakash Murali and Dripto M. Debroy and Kenneth R. Brown and Margaret Martonosi},
  title        = {Architecting {N}oisy {I}ntermediate{-S}cale {T}rapped {I}on {Q}uantum {C}omputers},
  booktitle    = {{ACM/IEEE} 47th Annu. Int. Symp. Comput. Archit. (ISCA)},
  pages        = {529--542},
  publisher    = {IEEE},
  address      = {Virtual Event},
  year         = {May 30--June 3, 2020},
  doi          = {10.1109/ISCA45697.2020.00051},
  url          = {https://www.doi.org/10.1109/ISCA45697.2020.00051},
  eprint       = {2004.04706},
}

@misc{NakajiWHCPKA25,
  author       = {Kouhei Nakaji and Jonathan Wurtz and Haozhe Huang and Luis Mantilla Calder{\'o}n and Karthik Panicker and Elica Kyoseva and Al{\'a}n Aspuru-Guzik},
  title        = {Quantum {C}ircuits as a {G}ame: {A} {R}einforcement {L}earning {A}gent for {Q}uantum {C}ompilation and its {A}pplication to {R}econfigurable {N}eutral {A}tom {A}rrays},
  publisher    = {arXiv},
  archivePrefix= {arXiv},
  primaryClass = {quant-ph},
  year         = {2025},
  month        = jun,
  doi          = {10.48550/arXiv.2506.05536},
  url          = {https://www.doi.org/10.48550/arXiv.2506.05536},
  eprint       = {2506.05536},
}

@misc{OpenAI25,
  author       = {{OpenAI}},
  title        = {gpt-oss-120b \& gpt-oss-20b {M}odel {C}ard},
  publisher    = {arXiv},
  archivePrefix= {arXiv},
  primaryClass = {cs.CL},
  year         = {2025},
  month        = aug,
  doi          = {10.48550/arXiv.2508.10925},
  url          = {https://www.doi.org/10.48550/arXiv.2508.10925},
  eprint       = {2508.10925},
}

@article{PalerSFA23,
  author       = {Alexandru Paler and Lucian Sasu and Adrian{-}C\u{a}t\u{a}lin Florea and R\u{a}zvan Andonie},
  title        = {Machine {L}earning {O}ptimization of {Q}uantum {C}ircuit {L}ayouts},
  journal      = {{ACM} Trans. Quantum Comput. (TQC)},
  volume       = {4},
  number       = {2},
  pages        = {12:1--12:25},
  year         = {2023},
  month        = feb,
  doi          = {10.1145/3565271},
  url          = {https://www.doi.org/10.1145/3565271},
  eprint       = {2007.14608},
}

@inproceedings{PascoalFA24,
  author       = {Gonçalo Pascoal and João Paulo Fernandes and Rui Abreu},
  title        = {Deep {R}einforcement {L}earning {S}trategies for {N}oise{-A}daptive {Q}ubit {R}outing},
  booktitle    = {{IEEE} Int. Conf. Quantum Softw. (QSW)},
  pages        = {146--156},
  publisher    = {IEEE},
  address      = {Shenzhen, China},
  year         = {July 7--13, 2024},
  doi          = {10.1109/QSW62656.2024.00030},
  url          = {https://www.doi.org/10.1109/QSW62656.2024.00030},
}

@article{PinoDFGMABFHMRN21,
  author       = {J. M. Pino and J. M. Dreiling and C. Figgatt and J. P. Gaebler and S. A. Moses and M. S. Allman and C. H. Baldwin and M. Foss-Feig and D. Hayes and K. Mayer and C. Ryan{-}Anderson and B. Neyenhuis},
  title        = {Demonstration of the trapped-ion quantum {CCD} computer architecture},
  journal      = {Nature},
  volume       = {592},
  number       = {7853},
  pages        = {209--213},
  year         = {2021},
  month        = apr,
  doi          = {10.1038/s41586-021-03318-4},
  url          = {https://www.doi.org/10.1038/s41586-021-03318-4},
  eprint       = {2003.01293},
}

@article{PozziHSM22,
  author       = {Matteo G. Pozzi and Steven J. Herbert and Akash Sengupta and Robert D. Mullins},
  title        = {Using {R}einforcement {L}earning to {P}erform {Q}ubit {R}outing in {Q}uantum {C}ompilers},
  journal      = {{ACM} Trans. Quantum Comput. (TQC)},
  volume       = {3},
  number       = {2},
  pages        = {10:1--10:25},
  year         = {2022},
  month        = may,
  doi          = {10.1145/3520434},
  url          = {https://www.doi.org/10.1145/3520434},
  eprint       = {2007.15957},
}

@inproceedings{PromponasMCPST25,
  author       = {Panagiotis Promponas and Akrit Mudvari and Luca Della Chiesa and Paul Polakos and Louis Samuel and Leandros Tassiulas},
  title        = {Compiler for {D}istributed {Q}uantum {C}omputing: {A} {R}einforcement {L}earning {A}pproach},
  booktitle    = {{IEEE} Int. Conf. Commun. (ICC)},
  pages        = {4615--4621},
  publisher    = {IEEE},
  address      = {Montreal, Canada},
  year         = {June 8--12, 2025},
  doi          = {10.1109/ICC52391.2025.11161597},
  url          = {https://www.doi.org/10.1109/ICC52391.2025.11161597},
  eprint       = {2404.17077},
}

@misc{QwenTeam25,
  author       = {{Qwen Team}},
  title        = {Qwen3 {T}echnical {R}eport},
  publisher    = {arXiv},
  archivePrefix= {arXiv},
  primaryClass = {cs.CL},
  year         = {2025},
  month        = may,
  doi          = {10.48550/arXiv.2505.09388},
  url          = {https://www.doi.org/10.48550/arXiv.2505.09388},
  eprint       = {2505.09388},
}

@inproceedings{RafailovSMEMF23,
  author        = {Rafael Rafailov and Archit Sharma and Eric Mitchell and Stefano Ermon and Christopher D. Manning and Chelsea Finn},
  title         = {Direct {P}reference {O}ptimization:
{Y}our {L}anguage {M}odel is {S}ecretly a {R}eward {M}odel},
  booktitle    = {37th Int. Conf. Neural Inf. Process. Syst. (NeurIPS)},
  pages        = {53728--53741},
  publisher    = {Curran Associates Inc.},
  address      = {New Orleans, LA, USA},
  year         = {Dec. 10--16, 2023},
  url          = {https://dl.acm.org/doi/10.5555/3666122.3668460},
  eprint       = {2305.18290},
}

@inproceedings{RajbhandariRRH20,
  author       = {Samyam Rajbhandari and Jeff Rasley and Olatunji Ruwase and Yuxiong He},
  title        = {Ze{RO}: Memory optimizations {T}oward {T}raining {T}rillion {P}arameter {M}odels},
  booktitle    = {Int. Conf. High Perform. Comput., Netw., Storage Anal. (SC20)},
  pages        = {20:1--20:16},
  publisher    = {IEEE},
  address      = {Atlanta, GA, USA},
  year         = {Nov. 9--19, 2020},
  doi          = {10.1109/SC41405.2020.00024},
  url          = {https://www.doi.org/10.1109/SC41405.2020.00024},
  eprint       = {1910.02054},
}

@article{Ransford+26,
  author       = {Anthony Ransford and others},
  title        = {A 98{-}qubit trapped-ion quantum computer with all{-}to{-}all connectivity},
  journal      = {Nature},
  volume       = {655},
  number       = {8121},
  pages        = {81--86},
  year         = {2026},
  month        = jul,
  doi          = {10.1038/s41586-026-10676-4},
  url          = {https://www.doi.org/10.1038/s41586-026-10676-4},
  eprint       = {2511.05465},
}

@inproceedings{RasleyRRH20,
  author       = {Jeff Rasley and Samyam Rajbhandari and Olatunji Ruwase and Yuxiong He},
  title        = {Deep{S}peed: {S}ystem {O}ptimizations {E}nable {T}raining {D}eep {L}earning {M}odels with {O}ver 100 {B}illion {P}arameters},
  booktitle    = {26th {ACM} {SIGKDD} Int. Conf. Knowl. Discovery \& Data Mining},
  pages        = {3505--3506},
  publisher    = {ACM},
  address      = {Virtual Event},
  year         = {July 6--10, 2020},
  doi          = {10.1145/3394486.3406703},
  url          = {https://www.doi.org/10.1145/3394486.3406703},
}

@inproceedings{RenRARYZLH21,
  author       = {Jie Ren and Samyam Rajbhandari and Reza Yazdani Aminabadi and Olatunji Ruwase and Shuangyan Yang and Minjia Zhang and Dong Li and Yuxiong He},
  title        = {Ze{RO-O}ffload: {D}emocratizing {B}illion{-S}cale {M}odel {T}raining},
  booktitle    = {USENIX Annu. Tech. Conf. (USENIX ATC 21)},
  pages        = {551--564},
  publisher    = {USENIX Association},
  address      = {Virtual event},
  year         = {July 14--16, 2021},
  url          = {https://www.usenix.org/conference/atc21/presentation/ren-jie},
  eprint       = {2101.06840},
}

@misc{RuanZFLCHHHHPD25,
  author       = {Jixuan Ruan and Hezi Zhang and Xiang Fang and Ang Li and Wesley C. Campbell and Eric Hudson and David Hayes and Hartmut Haeffner and Travis Humble and Jens Palsberg and Yufei Ding},
  title        = {{TrapSIMD}: {SIMD-A}ware {C}ompiler {O}ptimization for {2D} {T}rapped{-I}on {Q}uantum {M}achines},
  publisher    = {arXiv},
  archivePrefix= {arXiv},
  primaryClass = {quant-ph},
  year         = {2025},
  month        = apr,
  doi          = {10.48550/arXiv.2504.17886},
  url          = {https://www.doi.org/10.48550/arXiv.2504.17886},
  eprint       = {2504.17886},
}

@misc{RussonBYS26,
  author       = {Brent Russon and Bao G. Bach and Ed Younis and Ilya Safro},
  title        = {Scaling {Q}ubit {M}apping and {R}outing {W}ith {P}osition {G}raph {A}bstraction and {M}emoization},
  publisher    = {arXiv},
  archivePrefix= {arXiv},
  primaryClass = {quant-ph},
  year         = {2026},
  month        = may,
  doi          = {10.48550/arXiv.2605.09237},
  url          = {https://www.doi.org/10.48550/arXiv.2605.09237},
  eprint       = {2605.09237}
}

@inproceedings{RussoPPAC25,
  author       = {Enrico Russo and Maurizio Palesi and Davide Patti and Giuseppe Ascia and Vincenzo Catania},
  title        = {Optimizing {Q}ubit {A}ssignment in {M}odular {Q}uantum {S}ystems via {A}ttention{-B}ased {D}eep {R}einforcement {L}earning},
  booktitle    = {Des. Automat. Test Europe Conf. (DATE)},
  pages        = {1--7},
  publisher    = {IEEE},
  address      = {Lyon, France},
  year         = {Mar. 31--Apr. 2, 2025},
  doi          = {10.23919/DATE64628.2025.10992725},
  url          = {https://www.doi.org/10.23919/DATE64628.2025.10992725},
  eprint       = {2406.11452},
}

@article{SangHH26,
  author       = {Sokea Sang and Leanghok Hour and Youngsun Han},
  title        = {Learning{-}optimized qubit mapping and reuse to minimize intercore communication in modular quantum architectures},
  journal      = {Phys. Rev. Applied},
  volume       = {25},
  number       = {5},
  pages        = {054023},
  year         = {2026},
  month        = may,
  doi          = {10.1103/rfn7-c5n6},
  url          = {https://www.doi.org/10.1103/rfn7-c5n6},
  eprint       = {2506.09323},
}

@misc{SchierRSSBHR26,
  author       = {Maximilian Schier and Lea Richtmann and Christian Staufenbiel and Tobias Schmale and Daniel Borcherding and Mich{\`e}le Heurs and Bodo Rosenhahn},
  title        = {Reinforcement {L}earning for {I}on {S}huttling on {T}rapped-{I}on {Q}uantum {C}omputers},
  publisher    = {arXiv},
  archivePrefix= {arXiv},
  primaryClass = {quant-ph},
  year         = {2026},
  month        = may,
  doi          = {10.48550/arXiv.2605.22463},
  url          = {https://www.doi.org/10.48550/arXiv.2605.22463},
  eprint       = {2605.22463},
}

@inproceedings{SchmaleTBPKDOWB22,
  author       = {Tobias Schmale and Bence Temesi and Alakesh Baishya and Nicolas Pulido{-}Mateo and Ludwig Krinner and Timko Dubielzig and Christian Ospelkaus and Hendrik Weimer and Daniel Borcherding},
  title        = {Backend compiler phases for trapped-ion quantum computers},
  booktitle    = {{IEEE} Int. Conf. Quantum Softw. (QSW)},
  pages        = {32--37},
  publisher    = {IEEE},
  address      = {Barcelona, Spain},
  year         = {July 10--16, 2022},
  doi          = {10.1109/QSW55613.2022.00020},
  url          = {https://www.doi.org/10.1109/QSW55613.2022.00020},
  eprint       = {2206.00544},
}

@inproceedings{SchoenbergerHBW24,
  author       = {Daniel Schoenberger and Stefan Hillmich and Matthias Brandl and Robert Wille},
  title        = {Using {B}oolean {S}atisfiability for {E}xact {S}huttling in {T}rapped{-I}on {Q}uantum {C}omputers},
  booktitle    = {29th Asia South Pacific Des. Automat. Conf. (ASP-DAC)},
  pages        = {127--133},
  publisher    = {IEEE},
  address      = {Incheon, Republic of Korea},
  year         = {Jan. 22--25, 2024},
  doi          = {10.1109/ASP-DAC58780.2024.10473902},
  url          = {https://www.doi.org/10.1109/ASP-DAC58780.2024.10473902},
  eprint       = {2311.03454},
}

@article{SchoenbergerHBW25,
  author       = {Daniel Schoenberger and Stefan Hillmich and Matthias Brandl and Robert Wille},
  title        = {Shuttling for {S}calable {T}rapped{-I}on {Q}uantum {C}omputers},
  journal      = {{IEEE} Trans. Comp.-Aided Des. Integ. Cir. Sys.},
  volume       = {44},
  number       = {6},
  pages        = {2144--2155},
  year         = {2025},
  month        = jun,
  doi          = {10.1109/TCAD.2024.3513262},
  url          = {https://www.doi.org/10.1109/TCAD.2024.3513262},
  eprint       = {2402.14065},
}

@inproceedings{SchoenbergerHSW25,
  author       = {Daniel Schoenberger and Janine Hilder and Ferdinand Schmidt-Kaler and Robert Wille},
  title        = {Shuttling for {T}rapped{-I}on {Q}uantum {C}omputers with {E}mbedded {P}rocessing {Z}ones},
  booktitle    = {{IEEE} Int. Conf. Quantum Softw. (QSW)},
  pages        = {123--129},
  publisher    = {IEEE},
  address      = {Helsinki, Finland},
  year         = {July 7--12, 2025},
  doi          = {10.1109/QSW67625.2025.00023},
  url          = {https://www.doi.org/10.1109/QSW67625.2025.00023},
}

@inproceedings{SchoenbergerW25,
  author       = {Daniel Schoenberger and Robert Wille},
  title        = {Orchestrating {M}ulti{-Z}one {S}huttling in {T}rapped{-I}on {Q}uantum {C}omputers},
  booktitle    = {{IEEE} Int. Conf. Quantum Comput. Eng. (QCE)},
  volume       = {1},
  pages        = {1069--1075},
  publisher    = {IEEE},
  address      = {Albuquerque, NM, USA},
  year         = {Aug. 31--Sept. 5, 2025},
  doi          = {10.1109/QCE65121.2025.00119},
  url          = {https://www.doi.org/10.1109/QCE65121.2025.00119},
  eprint       = {2505.07928},
}

@article{SchulzPSS06,
  author       = {Stephan Schulz and Ulrich Poschinger and Kilian Singer and Ferdinand Schmidt{-}Kaler},
  title        = {Optimization of segmented linear {P}aul traps and transport of stored particles},
  journal      = {Fortschr. Phys.},
  volume       = {54},
  number       = {8--10},
  pages        = {648--665},
  year         = {2006},
  month        = aug,
  doi          = {10.1002/prop.200610324},
  url          = {https://www.doi.org/10.1002/prop.200610324},
  eprint       = {quant-ph/0607217},
}

@misc{ShaoWZXS24,
  author       = {Zhihong Shao and Peiyi Wang and Qihao Zhu and Runxin Xu and Junxiao Song and Xiao Bi and Haowei Zhang and Mingchuan Zhang and Y. K. Li and Y. Wu and Daya Guo},
  title        = {Deep{S}eek{M}ath: {P}ushing the {L}imits of {M}athematical {R}easoning in {O}pen {L}anguage {M}odels},
  publisher    = {arXiv},
  archivePrefix= {arXiv},
  primaryClass = {cs.CL},
  year         = {2024},
  month        = feb,
  doi          = {10.48550/arXiv.2402.03300},
  url          = {https://www.doi.org/10.48550/arXiv.2402.03300},
  eprint       = {2402.03300},
}

@article{ShuVBNVSB14,
  author       = {G. Shu and G. Vittorini and A. Buikema and C. S. Nichols and C. Volin and D. Stick and Kenneth R. Brown},
  title        = {Heating rates and ion{-}motion control in a $\mathsf{Y}${-}junction surface{-}electrode trap},
  journal      = {Phys. Rev. A},
  volume       = {89},
  number       = {6},
  pages        = {062308},
  year         = {2014},
  month        = jun,
  doi          = {10.1103/PhysRevA.89.062308},
  url          = {https://www.doi.org/10.1103/PhysRevA.89.062308},
}

@inproceedings{SinhaAS22,
  author       = {Animesh Sinha and Utkarsh Azad and Harjinder Singh},
  title        = {Qubit {R}outing {U}sing {G}raph {N}eural {N}etwork {A}ided {M}onte {C}arlo {T}ree {S}earch},
  booktitle    = {36th {AAAI} Conf. Artif. Intell.},
  pages        = {9935--9943},
  publisher    = {AAAI},
  address      = {Virtual Event},
  year         = {Feb. 22--Mar. 1, 2022},
  doi          = {10.1609/aaai.v36i9.21231},
  url          = {https://www.doi.org/10.1609/aaai.v36i9.21231},
  eprint       = {2104.01992},
}

@article{SivarajahDCSED20,
  author       = {Seyon Sivarajah and Silas Dilkes and Alexander Cowtan and Will Simmons and Alec Edgington and Ross Duncan},
  title        = {t$\vert$ket$\rangle$: a retargetable compiler for {NISQ} devices},
  journal      = {Quantum Sci. Technol.},
  volume       = {6},
  number       = {1},
  pages        = {014003},
  year         = {2020},
  month        = nov,
  doi          = {10.1088/2058-9565/ab8e92},
  url          = {https://www.doi.org/10.1088/2058-9565/ab8e92},
  eprint       = {2003.10611},
}

@inproceedings{TangDKFKS24,
  author       = {Wei Tang and Yiheng Duan and Yaroslav Kharkov and Rasool Fakoor and Eric Kessler and Yunong Shi},
  title        = {Alpha{R}outer: {Q}uantum {C}ircuit {R}outing with {R}einforcement {L}earning and {T}ree {S}earch},
  booktitle    = {{IEEE} Int. Conf. Quantum Comput. Eng. (QCE)},
  volume       = {1},
  pages        = {930--940},
  address      = {Montreal, Canada},
  publisher    = {IEEE},
  year         = {Sept. 15--20, 2024},
  doi          = {10.1109/QCE60285.2024.00112},
  url          = {https://www.doi.org/10.1109/QCE60285.2024.00112},
  eprint       = {2410.05115},
}

@repository{TaoriGZDLGLH23,
  author       = {Rohan Taori and Ishaan Gulrajani and Tianyi Zhang and Yann Dubois and Xuechen Li and Carlos Guestrin and Percy Liang and Tatsunori B. Hashimoto },
  title        = {Stanford {A}lpaca: {A}n {I}nstruction{-}following {LLaMA} model},
  year         = {2023},
  month        = may,
  note         = {{G}it{H}ub repository},
  code         = {https://github.com/tatsu-lab/stanford\_alpaca},
  commit       = {761dc5b},
}

@inproceedings{VelascoTOG25,
  author       = {Ander Artola Velasco and Stratis Tsirtsis and Nastaran Okati and Manuel Gomez{-}Rodriguez},
  title        = {Is {Y}our {LLM} {O}vercharging {Y}ou{?} {T}okenization{,} {T}ransparency{,} and {I}ncentives}, 
  booktitle    = {Tokenization Workshop (TokShop) @ 42nd Int. Conf. Mach. Learn. (ICML)},
  pages        = {1--21},
  publisher    = {OpenReview},
  address      = {Vancouver, Canada},
  year         = {July 18, 2025},
  url          = {https://openreview.net/forum?id=ht1vGnT3vF},
  eprint       = {2505.21627},
}

@mastersthesis{Wagner22,
  author       = {Janis Wagner},
  title        = {Automated generation of shuttling schedules for a scalable trapped-ion quantum processor},
  school       = {Institute of Computer Science, Johannes Gutenberg Univ.},
  address      = {Mainz, Germany},
  year         = {2022},
  month        = jul,
  type         = {M.Sc. thesis},
}

@article{WebberHWH20,
  author       = {Mark Webber and Steven Herbert and Sebastian Weidt and Winfried K. Hensinger},
  title        = {Efficient {Q}ubit {R}outing for a {G}lobally {C}onnected {T}rapped {I}on {Q}uantum {C}omputer},
  journal      = {Adv. Quantum Technol.},
  volume       = {3},
  number       = {8},
  pages        = {2000027},
  year         = {2020},
  month        = aug,
  doi          = {10.1002/qute.202000027},
  url          = {https://www.doi.org/10.1002/qute.202000027},
  eprint       = {2002.12782},
}

@inproceedings{Wolf+20,
  author       = {Thomas Wolf and others},
  title        = {Transformers: {S}tate{-}of{-}the{-A}rt {N}atural {L}anguage {P}rocessing},
  booktitle    = {Conf. Empirical Methods Natural Lang. Process.: Syst. Demonstrations},
  pages        = {38--45},
  publisher    = {ACL},
  address      = {Online Event},
  year         = {Nov. 16--20, 2020},
  doi          = {10.18653/v1/2020.emnlp-demos.6},
  url          = {https://www.doi.org/10.18653/v1/2020.emnlp-demos.6},
  eprint       = {1910.03771},
}

@article{WrightAFVDHPLDKSH13,
  author       = {Kenneth Wright and Jason M. Amini and Daniel L. Faircloth and Curtis Volin and S. Charles Doret and Harley Hayden and C.{-}S. Pai and David W. Landgren and Douglas Denison and Tyler Killian and Richart E. Slusher and Alexa W. Harter},
  title        = {Reliable transport through a microfabricated {X-}junction surface{-}electrode ion trap},
  journal      = {New J. Phys.},
  volume       = {15},
  number       = {3},
  pages        = {033004},
  year         = {2013},
  month        = mar,
  doi          = {10.1088/1367-2630/15/3/033004},
  url          = {https://www.doi.org/10.1088/1367-2630/15/3/033004},
  eprint       = {1210.3655},
}

@inproceedings{WuW26,
  author       = {Tung{-}Yeh Wu and Ting{-}Chi Wang},
  title        = {An {I}mproved {I}on{-S}huttling {A}pproach for {QCCD} {A}rchitectures},
  booktitle    = {Int. Symp. Phys. Des. (ISPD)},
  pages        = {173--181},
  publisher    = {ACM},
  address      = {Bonn, Germany},
  year         = {Mar. 15--18, 2026},
  doi          = {10.1145/3764386.3779585},
  url          = {https://www.doi.org/10.1145/3764386.3779585},
}

@inproceedings{WuZWW25,
  author       = {Xian Wu and Chenghong Zhu and Jingbo Wang and Xin Wang},
  title        = {{MUSS-TI}: {M}ulti{-}level {S}huttle {S}cheduling for {L}arge{-S}cale {E}ntanglement {M}odule {L}inked {T}rapped{-I}on},
  booktitle    = {58th {IEEE/ACM} Int. Symp. Microarchit. (MICRO)},
  pages        = {749--763},
  publisher    = {ACM},
  address      = {Seoul, Republic of Korea},
  year         = {Oct. 18--22, 2025},
  doi          = {10.1145/3725843.3756129},
  url          = {https://www.doi.org/10.1145/3725843.3756129},
  eprint       = {2509.25988},
}

@misc{ZeynaliB25,
  author       = {Atiye Zeynali and Zahra Bakhshi},
  title        = {Noise{-A}daptive {Q}uantum {C}ircuit {M}apping for {M}ulti{-C}hip {NISQ S}ystems via {D}eep {R}einforcement {L}earning},
  publisher    = {arXiv},
  archivePrefix= {arXiv},
  primaryClass = {quant-ph},  
  year         = {2025},
  month        = nov,
  doi          = {10.48550/arXiv.2511.18079},
  url          = {https://www.doi.org/10.48550/arXiv.2511.18079},
  eprint       = {2511.18079},
}

@article{Zhao+23,
  author       = {Yanli Zhao and others},
  title        = {Py{T}orch {FSDP}: {E}xperiences on {S}caling {F}ully {S}harded {D}ata {P}arallel},
  journal      = {Proc. {VLDB} Endow.},
  volume       = {16},
  number       = {12},
  pages        = {3848–-3860},
  year         = {2023},
  month        = aug,
  doi          = {10.14778/3611540.3611569},
  url          = {https://www.doi.org/10.14778/3611540.3611569},
  eprint       = {2304.11277},
}

@article{ZhaoZLT+26,
  author       = {Wayne Xin Zhao and Kun Zhou and Junyi Li and others},
  title        = {A {S}urvey of {L}arge {L}anguage {M}odels},
  journal      = {Front. Comput. Sci.},
  volume       = {20},
  number       = {12},
  pages        = {2012627},
  year         = {2026},
  month        = may,
  doi          = {10.1007/s11704-026-60308-3},
  url          = {https://www.doi.org/10.1007/s11704-026-60308-3},
  eprint       = {2303.18223},
}

@repository{Zhou19,
  author       = {Xiangzhen Zhou},
  year         = {2019},
  month        = aug,
  note         = {{G}it{H}ub repository},
  code         = {https://github.com/BensonZhou1991/circuittransform/tree/master/inputs/QASM\%20example},
  commit       = {8e6e5b1},
}

@article{ZhouFL22,
  author       = {Xiangzhen Zhou and Yuan Feng and Sanjiang Li},
  title        = {Supervised {L}earning {E}nhanced {Q}uantum {C}ircuit {T}ransformation},
  journal      = {{IEEE} Trans. Comput.-Aided Des. Integr. Circuits Syst. (TCAD)},
  volume       = {42},
  number       = {2},
  pages        = {437--447},
  year         = {2023},
  doi          = {10.1109/TCAD.2022.3179223},
  url          = {https://www.doi.org/10.1109/TCAD.2022.3179223},
  eprint       = {2110.03057},
}

@article{ZhouLF20,
  author       = {Xiangzhen Zhou and Sanjiang Li and Yuan Feng},
  title        = {Quantum {C}ircuit {T}ransformation {B}ased on {S}imulated {A}nnealing and {H}euristic {S}earch},
  journal      = {{IEEE} Trans. Comp.-Aided Des. Integ. Cir. Sys.},
  volume       = {39},
  number       = {12},
  pages        = {4683--4694},
  year         = {2020},
  month        = dec,
  doi          = {10.1109/TCAD.2020.2969647},
  url          = {https://www.doi.org/10.1109/TCAD.2020.2969647},
  eprint       = {1908.08853},
}

@inproceedings{ZhuWWW25,
  author       = {Chenghong Zhu and Xian Wu and Jingbo Wang and Xin Wang},
  title        = {{S}-{SYNC}: {S}huttle and {S}wap {C}o-{O}ptimization in {Q}uantum {C}harge-{C}oupled {D}evices},
  booktitle    = {52nd Annu. Int. Symp. Comput. Archit. (ISCA)},
  pages        = {271--284},
  publisher    = {ACM},
  address      = {Tokyo, Japan},
  year         = {June 21--25, 2025},
  doi          = {10.1145/3695053.3731084},
  url          = {https://www.doi.org/10.1145/3695053.3731084},
  eprint       = {2505.01316},
}

\newcommand{\dashrule}[1]{\noalign{\vskip1.2pt\hbox to \textwidth{\leaders\hbox to 3.5pt{\hss\vrule width2pt height0.4pt\hss}\hfill}\vskip1.2pt}}

\clearpage
\onecolumn
\appendix
\appendixpage
\addappheadtotoc
\section{Complete results}
\label{sec:appendix}

\noindent This appendix lists every configuration of LLM, sampling temperature, circuit, and trained architecture that produced at least one complete, valid schedule in its ten runs, together with the operation counts, retries, response tokens, and runtimes summarized in \autoref{subsec:results}. \autoref{tab:linear} covers the linear architecture, \autoref{tab:branched-a} the branched architecture for \emph{4mod5-bdd\_287} with Llama-3B, Qwen-4B, and Qwen-14B, and \autoref{tab:branched-b} the remaining LLMs on that circuit together with \emph{mini\_alu\_305} and \emph{cnt3-5\_179}.

\begin{table}[!ht]
	\centering
	\scriptsize
	\setlength{\tabcolsep}{1.6pt}
	\caption{Configurations on the linear architecture that produced at least one complete, valid schedule in ten runs. \emph{Runs} counts the runs that completed. \emph{Base} is the operation count of the baseline \cite{Wagner22, KreppelMWHPSB24} for that configuration, \emph{Best} the smallest count over the ten runs, and $\Delta$ the deviation of \emph{Best} from \emph{Base}. \emph{Mean\,$\pm$\,SD} is the mean and standard deviation (SD) over the completed runs. \emph{Raw} is the best run counted before post-processing and $\Delta_\text{raw}$ its deviation from \emph{Base}. \emph{Retr.} is the number of retries over the best run and \emph{Tokens} its response tokens without and with retries. \emph{Time} is the mean runtime over the completed runs. Negative $\Delta$ is set in bold.}
	\label{tab:linear}
	\begin{tabular*}{\textwidth}{@{\extracolsep{\fill}}lcccrrrcrrrr@{}}
		\toprule
		LLM & $T$ & Runs & Base & Best & $\Delta$ & Mean $\pm$ SD & Raw & $\Delta_\text{raw}$ & Retr. & Tokens [k] & Time [min] \\
		\midrule[1.2pt]
		\multicolumn{12}{c}{\textbf{Circuit: 4mod5-bdd\_287 (7 qubits, 106 gates)}} \\
		\midrule
		Llama-3B & 0.7 & 8/10 & 607 & 559 & \textbf{-7.9} & 670 $\pm$ 87 & 561 & \textbf{-7.6} & 9 & 61\,/\,183 & 29 \\
		Llama-3B & 1.0 & 8/10 & 607 & 653 & +7.6 & 812 $\pm$ 135 & 665 & +9.6 & 9 & 78\,/\,283 & 56 \\
		Qwen-4B & 0.7 & 1/10 & 607 & 1{,}071 & +76.4 & 1{,}071 $\pm$ -- & 1{,}127 & +85.7 & 19 & 143\,/\,465 & 81 \\
		Qwen-4B & 1.0 & 4/10 & 607 & 517 & \textbf{-14.8} & 741 $\pm$ 231 & 529 & \textbf{-12.9} & 26 & 60\,/\,372 & 41 \\
		Qwen-14B & 0.7 & 10/10 & 607 & 842 & +38.7 & 870 $\pm$ 37 & 844 & +39.0 & 2 & 100\,/\,180 & 93 \\
		Qwen-14B & 1.0 & 10/10 & 607 & 701 & +15.5 & 871 $\pm$ 111 & 721 & +18.8 & 17 & 86\,/\,416 & 185 \\
		DeepSeek-7B & 0.7 & 4/10 & 607 & 598 & \textbf{-1.5} & 641 $\pm$ 63 & 598 & \textbf{-1.5} & 5 & 76\,/\,102 & 37 \\
		DeepSeek-7B & 1.0 & 8/10 & 607 & 543 & \textbf{-10.5} & 671 $\pm$ 107 & 557 & \textbf{-8.2} & 15 & 73\,/\,172 & 52 \\
		Gemma-27B & 0.7 & 6/10 & 607 & 700 & +15.3 & 881 $\pm$ 123 & 714 & +17.6 & 17 & 86\,/\,250 & 143 \\
		Gemma-27B & 1.0 & 2/10 & 607 & 825 & +35.9 & 878 $\pm$ 76 & 829 & +36.6 & 15 & 104\,/\,272 & 149 \\
		\midrule[1.2pt]
		\multicolumn{12}{c}{\textbf{Circuit: mini\_alu\_305 (10 qubits, 263 gates)}} \\
		\midrule
		Qwen-4B & 1.0 & 1/10 & 2{,}390 & 3{,}124 & +30.7 & 3{,}124 $\pm$ -- & 3{,}200 & +33.9 & 33 & 555\,/\,1{,}252 & 211 \\
		\bottomrule
	\end{tabular*}
\end{table}

\begin{table}[!ht]
	\centering
	\scriptsize
	\setlength{\tabcolsep}{1.6pt}
	\caption{Configurations on the branched architecture that produced at least one complete, valid schedule in ten runs, listed for every combination of stack height $s$ and junction distance $d$ that succeeded. Columns are as in \autoref{tab:linear}.}
	\label{tab:branched-a}
	\begin{tabular*}{\textwidth}{@{\extracolsep{\fill}}lccccrrrcrrrr@{}}
		\toprule
		LLM & $T$ & $(s,d)$ & Runs & Base & Best & $\Delta$ & Mean $\pm$ SD & Raw & $\Delta_\text{raw}$ & Retr. & Tokens [k] & Time [min] \\
		\midrule[1.2pt]
		\multicolumn{13}{c}{\textbf{Circuit: 4mod5-bdd\_287 (7 qubits, 106 gates)}} \\
		\midrule
		Llama-3B & 0.7 & (1,\,1) & 10/10 & 493 & 528 & +7.1 & 631 $\pm$ 60 & 538 & +9.1 & 0 & 67\,/\,67 & 11 \\
		 &  & (1,\,2) & 10/10 & 484 & 554 & +14.5 & 632 $\pm$ 51 & 614 & +26.9 & 2 & 77\,/\,85 & 16 \\
		 &  & (1,\,3) & 6/10 & 482 & 523 & +8.5 & 564 $\pm$ 37 & 617 & +28.0 & 6 & 81\,/\,152 & 21 \\
		 &  & (1,\,4) & 6/10 & 484 & 508 & +5.0 & 549 $\pm$ 24 & 612 & +26.4 & 6 & 78\,/\,121 & 19 \\
		 &  & (2,\,1) & 7/10 & 503 & 506 & +0.6 & 564 $\pm$ 41 & 540 & +7.4 & 3 & 66\,/\,74 & 14 \\
		 &  & (2,\,2) & 8/10 & 496 & 525 & +5.8 & 573 $\pm$ 40 & 575 & +15.9 & 3 & 69\,/\,84 & 15 \\
		 &  & (2,\,3) & 7/10 & 489 & 524 & +7.2 & 626 $\pm$ 65 & 602 & +23.1 & 3 & 72\,/\,89 & 21 \\
		 &  & (3,\,1) & 10/10 & 520 & 561 & +7.9 & 639 $\pm$ 47 & 667 & +28.3 & 5 & 84\,/\,124 & 22 \\
		 &  & (4,\,1) & 7/10 & 515 & 532 & +3.3 & 592 $\pm$ 55 & 662 & +28.5 & 14 & 82\,/\,229 & 30 \\
		 &  & (5,\,1) & 3/10 & 694 & 566 & \textbf{-18.4} & 616 $\pm$ 86 & 698 & +0.6 & 10 & 83\,/\,300 & 44 \\
		\dashrule{13}
		Llama-3B & 1.0 & (1,\,1) & 9/10 & 493 & 582 & +18.1 & 680 $\pm$ 72 & 608 & +23.3 & 6 & 75\,/\,92 & 15 \\
		 &  & (1,\,2) & 10/10 & 484 & 573 & +18.4 & 655 $\pm$ 50 & 749 & +54.8 & 6 & 95\,/\,133 & 20 \\
		 &  & (1,\,3) & 9/10 & 482 & 522 & +8.3 & 576 $\pm$ 33 & 650 & +34.9 & 3 & 86\,/\,95 & 23 \\
		 &  & (1,\,4) & 9/10 & 484 & 484 & +0.0 & 569 $\pm$ 57 & 550 & +13.6 & 3 & 67\,/\,89 & 17 \\
		 &  & (2,\,1) & 6/10 & 503 & 536 & +6.6 & 577 $\pm$ 54 & 574 & +14.1 & 4 & 71\,/\,81 & 14 \\
		 &  & (2,\,2) & 7/10 & 496 & 534 & +7.7 & 609 $\pm$ 41 & 616 & +24.2 & 6 & 78\,/\,88 & 18 \\
		 &  & (2,\,3) & 9/10 & 489 & 464 & \textbf{-5.1} & 596 $\pm$ 75 & 498 & +1.8 & 1 & 56\,/\,57 & 17 \\
		 &  & (3,\,1) & 8/10 & 520 & 508 & \textbf{-2.3} & 593 $\pm$ 86 & 594 & +14.2 & 9 & 74\,/\,161 & 19 \\
		 &  & (4,\,1) & 8/10 & 515 & 546 & +6.0 & 697 $\pm$ 90 & 620 & +20.4 & 11 & 74\,/\,163 & 40 \\
		 &  & (5,\,1) & 1/10 & 694 & 615 & \textbf{-11.4} & 615 $\pm$ -- & 749 & +7.9 & 16 & 92\,/\,385 & 52 \\
		\dashrule{13}
		Qwen-4B & 0.7 & (1,\,1) & 1/10 & 493 & 592 & +20.1 & 592 $\pm$ -- & 622 & +26.2 & 29 & 80\,/\,176 & 27 \\
		 &  & (1,\,2) & 3/10 & 484 & 563 & +16.3 & 577 $\pm$ 13 & 623 & +28.7 & 17 & 86\,/\,158 & 23 \\
		 &  & (1,\,3) & 4/10 & 482 & 603 & +25.1 & 624 $\pm$ 33 & 661 & +37.1 & 14 & 86\,/\,127 & 36 \\
		 &  & (1,\,4) & 1/10 & 484 & 555 & +14.7 & 555 $\pm$ -- & 609 & +25.8 & 11 & 80\,/\,219 & 36 \\
		 &  & (2,\,2) & 6/10 & 496 & 547 & +10.3 & 590 $\pm$ 35 & 585 & +17.9 & 17 & 79\,/\,238 & 31 \\
		 &  & (2,\,3) & 3/10 & 489 & 589 & +20.4 & 623 $\pm$ 30 & 749 & +53.2 & 13 & 97\,/\,200 & 45 \\
		 &  & (3,\,1) & 5/10 & 520 & 600 & +15.4 & 720 $\pm$ 68 & 740 & +42.3 & 21 & 97\,/\,292 & 43 \\
		 &  & (5,\,1) & 2/10 & 694 & 704 & +1.4 & 724 $\pm$ 29 & 934 & +34.6 & 28 & 115\,/\,442 & 95 \\
		\dashrule{13}
		Qwen-4B & 1.0 & (1,\,1) & 3/10 & 493 & 558 & +13.2 & 628 $\pm$ 61 & 592 & +20.1 & 21 & 77\,/\,169 & 40 \\
		 &  & (1,\,2) & 9/10 & 484 & 537 & +11.0 & 598 $\pm$ 35 & 583 & +20.5 & 15 & 79\,/\,227 & 29 \\
		 &  & (1,\,3) & 4/10 & 482 & 602 & +24.9 & 636 $\pm$ 59 & 656 & +36.1 & 12 & 92\,/\,146 & 28 \\
		 &  & (1,\,4) & 3/10 & 484 & 634 & +31.0 & 674 $\pm$ 36 & 768 & +58.7 & 36 & 108\,/\,394 & 52 \\
		 &  & (2,\,2) & 6/10 & 496 & 540 & +8.9 & 596 $\pm$ 46 & 634 & +27.8 & 31 & 88\,/\,386 & 34 \\
		 &  & (2,\,3) & 6/10 & 489 & 576 & +17.8 & 701 $\pm$ 106 & 714 & +46.0 & 12 & 96\,/\,226 & 58 \\
		 &  & (3,\,1) & 7/10 & 520 & 580 & +11.5 & 660 $\pm$ 52 & 646 & +24.2 & 8 & 86\,/\,121 & 37 \\
		 &  & (5,\,1) & 2/10 & 694 & 620 & \textbf{-10.7} & 624 $\pm$ 6 & 910 & +31.1 & 31 & 113\,/\,499 & 95 \\
		\dashrule{13}
		Qwen-14B & 0.7 & (1,\,1) & 7/10 & 493 & 617 & +25.2 & 644 $\pm$ 27 & 651 & +32.0 & 9 & 87\,/\,159 & 96 \\
		 &  & (1,\,2) & 8/10 & 484 & 569 & +17.6 & 612 $\pm$ 33 & 659 & +36.2 & 27 & 88\,/\,308 & 113 \\
		 &  & (1,\,3) & 2/10 & 482 & 645 & +33.8 & 676 $\pm$ 45 & 837 & +73.7 & 20 & 118\,/\,395 & 194 \\
		 &  & (2,\,1) & 5/10 & 503 & 554 & +10.1 & 576 $\pm$ 24 & 626 & +24.5 & 29 & 87\,/\,445 & 143 \\
		 &  & (2,\,2) & 6/10 & 496 & 518 & +4.4 & 600 $\pm$ 41 & 680 & +37.1 & 22 & 98\,/\,365 & 139 \\
		 &  & (2,\,3) & 9/10 & 489 & 564 & +15.3 & 638 $\pm$ 73 & 616 & +26.0 & 15 & 75\,/\,198 & 159 \\
		 &  & (3,\,1) & 6/10 & 520 & 608 & +16.9 & 664 $\pm$ 47 & 694 & +33.5 & 11 & 95\,/\,223 & 121 \\
		 &  & (4,\,1) & 2/10 & 515 & 733 & +42.3 & 762 $\pm$ 41 & 921 & +78.8 & 29 & 122\,/\,764 & 312 \\
		 &  & (5,\,1) & 2/10 & 694 & 644 & \textbf{-7.2} & 670 $\pm$ 37 & 844 & +21.6 & 17 & 106\,/\,443 & 200 \\
		\dashrule{13}
		Qwen-14B & 1.0 & (1,\,1) & 6/10 & 493 & 572 & +16.0 & 619 $\pm$ 42 & 652 & +32.3 & 40 & 85\,/\,471 & 128 \\
		 &  & (1,\,2) & 7/10 & 484 & 566 & +16.9 & 631 $\pm$ 44 & 646 & +33.5 & 18 & 88\,/\,265 & 148 \\
		 &  & (1,\,3) & 2/10 & 482 & 658 & +36.5 & 682 $\pm$ 33 & 814 & +68.9 & 13 & 117\,/\,315 & 174 \\
		 &  & (2,\,1) & 3/10 & 503 & 612 & +21.7 & 627 $\pm$ 16 & 674 & +34.0 & 51 & 97\,/\,861 & 222 \\
		 &  & (2,\,2) & 8/10 & 496 & 545 & +9.9 & 611 $\pm$ 42 & 619 & +24.8 & 14 & 81\,/\,177 & 163 \\
		 &  & (2,\,3) & 7/10 & 489 & 553 & +13.1 & 627 $\pm$ 64 & 645 & +31.9 & 17 & 80\,/\,369 & 184 \\
		 &  & (3,\,1) & 6/10 & 520 & 592 & +13.8 & 703 $\pm$ 62 & 788 & +51.5 & 28 & 109\,/\,602 & 181 \\
		 &  & (4,\,1) & 4/10 & 515 & 591 & +14.8 & 634 $\pm$ 48 & 707 & +37.3 & 24 & 87\,/\,562 & 266 \\
		 &  & (5,\,1) & 1/10 & 694 & 550 & \textbf{-20.7} & 550 $\pm$ -- & 782 & +12.7 & 23 & 97\,/\,667 & 281 \\
		\bottomrule
	\end{tabular*}
\end{table}

\begin{table}[!ht]
	\centering
	\scriptsize
	\setlength{\tabcolsep}{1.6pt}
	\caption{Continuation of \autoref{tab:branched-a} for the remaining LLMs on 4mod5-bdd\_287 and for mini\_alu\_305 and cnt3-5\_179.}
	\label{tab:branched-b}
	\begin{tabular*}{\textwidth}{@{\extracolsep{\fill}}lccccrrrcrrrr@{}}
		\toprule
		LLM & $T$ & $(s,d)$ & Runs & Base & Best & $\Delta$ & Mean $\pm$ SD & Raw & $\Delta_\text{raw}$ & Retr. & Tokens [k] & Time [min] \\
		\midrule[1.2pt]
		\multicolumn{13}{c}{\textbf{Circuit: 4mod5-bdd\_287 (7 qubits, 106 gates)}} \\
		\midrule
		DeepSeek-7B & 0.7 & (3,\,1) & 2/10 & 520 & 541 & +4.0 & 582 $\pm$ 59 & 649 & +24.8 & 28 & 91\,/\,463 & 165 \\
		\dashrule{13}
		DeepSeek-7B & 1.0 & (2,\,1) & 1/10 & 503 & 566 & +12.5 & 566 $\pm$ -- & 624 & +24.1 & 40 & 94\,/\,389 & 107 \\
		 &  & (2,\,2) & 1/10 & 496 & 568 & +14.5 & 568 $\pm$ -- & 670 & +35.1 & 46 & 99\,/\,527 & 160 \\
		 &  & (2,\,3) & 2/10 & 489 & 639 & +30.7 & 654 $\pm$ 21 & 791 & +61.8 & 27 & 112\,/\,276 & 107 \\
		 &  & (3,\,1) & 1/10 & 520 & 719 & +38.3 & 719 $\pm$ -- & 841 & +61.7 & 47 & 122\,/\,564 & 169 \\
		 &  & (4,\,1) & 1/10 & 515 & 608 & +18.1 & 608 $\pm$ -- & 710 & +37.9 & 50 & 97\,/\,421 & 116 \\
		\dashrule{13}
		Gemma-27B & 0.7 & (1,\,2) & 1/10 & 484 & 654 & +35.1 & 654 $\pm$ -- & 694 & +43.4 & 39 & 102\,/\,446 & 289 \\
		 &  & (1,\,3) & 3/10 & 482 & 619 & +28.4 & 651 $\pm$ 31 & 679 & +40.9 & 23 & 93\,/\,266 & 228 \\
		 &  & (1,\,4) & 1/10 & 484 & 640 & +32.2 & 640 $\pm$ -- & 734 & +51.7 & 25 & 107\,/\,317 & 206 \\
		 &  & (2,\,3) & 3/10 & 489 & 533 & +9.0 & 583 $\pm$ 46 & 567 & +16.0 & 9 & 77\,/\,169 & 142 \\
		 &  & (4,\,1) & 1/10 & 515 & 571 & +10.9 & 571 $\pm$ -- & 649 & +26.0 & 10 & 85\,/\,145 & 94 \\
		\dashrule{13}
		Gemma-27B & 1.0 & (1,\,2) & 1/10 & 484 & 576 & +19.0 & 576 $\pm$ -- & 650 & +34.3 & 34 & 92\,/\,289 & 187 \\
		 &  & (1,\,3) & 4/10 & 482 & 543 & +12.7 & 616 $\pm$ 68 & 573 & +18.9 & 22 & 78\,/\,361 & 236 \\
		 &  & (1,\,4) & 1/10 & 484 & 671 & +38.6 & 671 $\pm$ -- & 723 & +49.4 & 20 & 101\,/\,235 & 152 \\
		 &  & (2,\,2) & 2/10 & 496 & 521 & +5.0 & 556 $\pm$ 50 & 555 & +11.9 & 25 & 74\,/\,239 & 210 \\
		 &  & (2,\,3) & 6/10 & 489 & 562 & +14.9 & 594 $\pm$ 28 & 642 & +31.3 & 15 & 86\,/\,218 & 168 \\
		 &  & (3,\,1) & 1/10 & 520 & 581 & +11.7 & 581 $\pm$ -- & 659 & +26.7 & 19 & 90\,/\,312 & 203 \\
		 &  & (4,\,1) & 4/10 & 515 & 560 & +8.7 & 622 $\pm$ 43 & 590 & +14.6 & 6 & 75\,/\,129 & 181 \\
		\midrule[1.2pt]
		\multicolumn{13}{c}{\textbf{Circuit: mini\_alu\_305 (10 qubits, 263 gates)}} \\
		\midrule
		Llama-3B & 0.7 & (1,\,1) & 10/10 & 1{,}660 & 1{,}921 & +15.7 & 2{,}086 $\pm$ 86 & 2{,}005 & +20.8 & 18 & 349\,/\,525 & 71 \\
		 &  & (1,\,2) & 6/10 & 1{,}769 & 1{,}897 & +7.2 & 2{,}035 $\pm$ 114 & 2{,}219 & +25.4 & 15 & 413\,/\,686 & 108 \\
		 &  & (1,\,3) & 1/10 & 1{,}847 & 2{,}160 & +16.9 & 2{,}160 $\pm$ -- & 2{,}648 & +43.4 & 47 & 493\,/\,1{,}402 & 188 \\
		 &  & (2,\,1) & 1/10 & 1{,}624 & 1{,}939 & +19.4 & 1{,}939 $\pm$ -- & 2{,}129 & +31.1 & 27 & 380\,/\,599 & 76 \\
		 &  & (2,\,2) & 2/10 & 1{,}706 & 1{,}932 & +13.2 & 1{,}982 $\pm$ 70 & 2{,}286 & +34.0 & 33 & 411\,/\,872 & 103 \\
		 &  & (2,\,3) & 6/10 & 1{,}864 & 1{,}791 & \textbf{-3.9} & 2{,}004 $\pm$ 112 & 2{,}269 & +21.7 & 26 & 401\,/\,726 & 123 \\
		 &  & (2,\,4) & 4/10 & 1{,}787 & 2{,}112 & +18.2 & 2{,}162 $\pm$ 53 & 2{,}746 & +53.7 & 65 & 498\,/\,1{,}443 & 158 \\
		 &  & (3,\,1) & 1/10 & 1{,}677 & 1{,}975 & +17.8 & 1{,}975 $\pm$ -- & 2{,}295 & +36.9 & 68 & 397\,/\,1{,}189 & 157 \\
		 &  & (3,\,2) & 1/10 & 1{,}726 & 1{,}898 & +10.0 & 1{,}898 $\pm$ -- & 2{,}328 & +34.9 & 29 & 412\,/\,843 & 110 \\
		 &  & (3,\,3) & 2/10 & 1{,}806 & 1{,}936 & +7.2 & 2{,}041 $\pm$ 148 & 2{,}436 & +34.9 & 40 & 421\,/\,1{,}086 & 150 \\
		 &  & (3,\,4) & 2/10 & 1{,}780 & 2{,}175 & +22.2 & 2{,}184 $\pm$ 13 & 2{,}921 & +64.1 & 48 & 492\,/\,1{,}436 & 173 \\
		 &  & (4,\,1) & 2/10 & 1{,}777 & 2{,}191 & +23.3 & 2{,}230 $\pm$ 55 & 2{,}515 & +41.5 & 80 & 418\,/\,1{,}727 & 265 \\
		 &  & (4,\,2) & 2/10 & 1{,}835 & 2{,}145 & +16.9 & 2{,}266 $\pm$ 172 & 2{,}615 & +42.5 & 25 & 431\,/\,863 & 145 \\
		 &  & (5,\,1) & 1/10 & 1{,}876 & 2{,}288 & +22.0 & 2{,}288 $\pm$ -- & 3{,}116 & +66.1 & 37 & 526\,/\,1{,}107 & 148 \\
		 &  & (6,\,1) & 1/10 & 1{,}890 & 2{,}682 & +41.9 & 2{,}682 $\pm$ -- & 3{,}254 & +72.2 & 58 & 525\,/\,1{,}668 & 229 \\
		\dashrule{13}
		Llama-3B & 1.0 & (1,\,1) & 10/10 & 1{,}660 & 2{,}048 & +23.4 & 2{,}179 $\pm$ 75 & 2{,}220 & +33.7 & 31 & 389\,/\,764 & 100 \\
		 &  & (1,\,2) & 6/10 & 1{,}769 & 2{,}077 & +17.4 & 2{,}116 $\pm$ 36 & 2{,}505 & +41.6 & 30 & 466\,/\,943 & 157 \\
		 &  & (2,\,1) & 2/10 & 1{,}624 & 2{,}002 & +23.3 & 2{,}026 $\pm$ 33 & 2{,}156 & +32.8 & 46 & 382\,/\,674 & 92 \\
		 &  & (2,\,2) & 9/10 & 1{,}706 & 1{,}809 & +6.0 & 2{,}024 $\pm$ 102 & 2{,}137 & +25.3 & 29 & 392\,/\,767 & 146 \\
		 &  & (2,\,3) & 5/10 & 1{,}864 & 1{,}837 & \textbf{-1.4} & 2{,}022 $\pm$ 122 & 2{,}239 & +20.1 & 31 & 392\,/\,888 & 155 \\
		 &  & (3,\,1) & 1/10 & 1{,}677 & 2{,}031 & +21.1 & 2{,}031 $\pm$ -- & 2{,}341 & +39.6 & 58 & 397\,/\,1{,}075 & 141 \\
		 &  & (3,\,3) & 3/10 & 1{,}806 & 1{,}730 & \textbf{-4.2} & 1{,}848 $\pm$ 107 & 2{,}248 & +24.5 & 53 & 396\,/\,1{,}485 & 198 \\
		 &  & (3,\,4) & 4/10 & 1{,}780 & 1{,}945 & +9.3 & 2{,}109 $\pm$ 125 & 2{,}417 & +35.8 & 64 & 403\,/\,1{,}644 & 199 \\
		 &  & (4,\,2) & 4/10 & 1{,}835 & 2{,}119 & +15.5 & 2{,}165 $\pm$ 75 & 2{,}685 & +46.3 & 57 & 443\,/\,1{,}484 & 210 \\
		 &  & (5,\,1) & 3/10 & 1{,}876 & 2{,}087 & +11.2 & 2{,}261 $\pm$ 214 & 2{,}649 & +41.2 & 54 & 440\,/\,1{,}363 & 193 \\
		 &  & (6,\,1) & 2/10 & 1{,}890 & 2{,}425 & +28.3 & 2{,}520 $\pm$ 134 & 3{,}167 & +67.6 & 61 & 512\,/\,1{,}723 & 264 \\
		\dashrule{13}
		Qwen-4B & 1.0 & (5,\,1) & 1/10 & 1{,}876 & 2{,}776 & +48.0 & 2{,}776 $\pm$ -- & 3{,}440 & +83.4 & 58 & 629\,/\,2{,}188 & 381 \\
		\midrule[1.2pt]
		\multicolumn{13}{c}{\textbf{Circuit: cnt3-5\_179 (16 qubits, 230 gates)}} \\
		\midrule
		Llama-3B & 0.7 & (1,\,1) & 5/10 & 1{,}935 & 2{,}435 & +25.8 & 2{,}575 $\pm$ 121 & 2{,}639 & +36.4 & 27 & 647\,/\,1{,}416 & 246 \\
		\dashrule{13}
		Llama-3B & 1.0 & (1,\,1) & 4/10 & 1{,}935 & 2{,}543 & +31.4 & 2{,}632 $\pm$ 66 & 2{,}655 & +37.2 & 65 & 654\,/\,2{,}261 & 297 \\
		 &  & (3,\,1) & 1/10 & 1{,}741 & 2{,}017 & +15.9 & 2{,}017 $\pm$ -- & 2{,}359 & +35.5 & 104 & 564\,/\,2{,}592 & 357 \\
		\bottomrule
	\end{tabular*}
\end{table}
\twocolumn

\end{document}